\documentclass{sig-alternate-2013-arxiv}
\usepackage{dsfont}
\usepackage{color}
\usepackage[ruled,linesnumbered]{algorithm2e}

\newfont{\mycrnotice}{ptmr8t at 7pt}
\newfont{\myconfname}{ptmri8t at 7pt}
\let\crnotice\mycrnotice%
\let\confname\myconfname%

\toappear{*Notice: To appear in the 24th ACM International Conference on Information and Knowledge Management (CIKM 2015).}
\begin{document}
%
% --- Author Metadata here ---
%\conferenceinfo{WOODSTOCK}{'97 El Paso, Texas USA}
%\CopyrightYear{2007} % Allows default copyright year (20XX) to be over-ridden - IF NEED BE.
%\crdata{0-12345-67-8/90/01}  % Allows default copyright data (0-89791-88-6/97/05) to be over-ridden - IF NEED BE.
% --- End of Author Metadata ---

\title{Cross-Modal Similarity Learning : A Low Rank \\Bilinear Formulation
%\titlenote{(Produces the permission block, and copyright information). For use with SIG-ALTERNATE.CLS. Supported by ACM. }
}

%\subtitle{[Extended Abstract]
%\titlenote{A full version of this paper is available as
%\textit{Author's Guide to Preparing ACM SIG Proceedings Using
%\LaTeX$2_\epsilon$\ and BibTeX} at
%\texttt{www.acm.org/eaddress.htm}}}
%
% You need the command \numberofauthors to handle the 'placement
% and alignment' of the authors beneath the title.
%
% For aesthetic reasons, we recommend 'three authors at a time'
% i.e. three 'name/affiliation blocks' be placed beneath the title.
%
% NOTE: You are NOT restricted in how many 'rows' of
% "name/affiliations" may appear. We just ask that you restrict
% the number of 'columns' to three.
%
% Because of the available 'opening page real-estate'
% we ask you to refrain from putting more than six authors
% (two rows with three columns) beneath the article title.
% More than six makes the first-page appear very cluttered indeed.
%
% Use the \alignauthor commands to handle the names
% and affiliations for an 'aesthetic maximum' of six authors.
% Add names, affiliations, addresses for
% the seventh etc. author(s) as the argument for the
% \additionalauthors command.
% These 'additional authors' will be output/set for you
% without further effort on your part as the last section in
% the body of your article BEFORE References or any Appendices.

%%----------------------------%%%------------------------------%%%
\numberofauthors{7} %  in this sample file, there are a *total*
%% of EIGHT authors. SIX appear on the 'first-page' (for formatting
%% reasons) and the remaining two appear in the \additionalauthors section.
%%
\author{
% You can go ahead and credit any number of authors here,
% e.g. one 'row of three' or two rows (consisting of one row of three
% and a second row of one, two or three).
%
% The command \alignauthor (no curly braces needed) should
% precede each author name, affiliation/snail-mail address and
% e-mail address. Additionally, tag each line of
% affiliation/address with \affaddr, and tag the
% e-mail address with \email.
%
% 1st. author
\alignauthor
Cuicui Kang%\titlenote{Corresponding author.}\\
\\
       \affaddr{Institute of Information Engineering, Chinese Academy of Sciences}\\
       \email{kangcuicui@iie.ac.cn}
% 2nd. author
\alignauthor
Shengcai Liao\\
       \affaddr{Institute of Automation, Chinese Academy of Sciences}\\
       \email{scliao@nlpr.ia.ac.cn}
% 3rd. author
\alignauthor
Yonghao He\\
       \affaddr{Institute of Automation, Chinese Academy of Sciences}\\
       \email{yhhe@nlpr.ia.ac.cn}
\and  % use '\and' if you need 'another row' of author names
% 4th. author
\alignauthor Jian Wang\\
       \affaddr{Institute of Automation, Chinese Academy of Sciences}\\
       \email{jian.wang@nlpr.ia.ac.cn}
% 5th. author
\alignauthor Wenjia Niu\\
       \affaddr{Institute of Information Engineering, Chinese Academy of Sciences}\\
       \email{niuwenjia@iie.ac.cn}
% 6th. author
\alignauthor Shiming Xiang\\
       \affaddr{Institute of Automation, Chinese Academy of Sciences}\\
       \email{smxiang@nlpr.ia.ac.cn}
%\alignauthor \\
%       \affaddr{Institute of Automation, Chinese Academy of Sciences}\\
%       \email{chpan@nlpr.ia.ac.cn}
}
%% There's nothing stopping you putting the seventh, eighth, etc.
%% author on the opening page (as the 'third row') but we ask,
%% for aesthetic reasons that you place these 'additional authors'
%% in the \additional authors block, viz.
\additionalauthors{Additional author: Chunhong Pan (Institute of Automation, Chinese Academy of Sciences,
email: {\texttt{chpan@nlpr.ia.ac.cn}}).}
%\date{30 July 1999}
%% Just remember to make sure that the TOTAL number of authors
%% is the number that will appear on the first page PLUS the
%% number that will appear in the \additionalauthors section.
%%%%%%%%%%%%%%---------------------------%%%%%%%%%%%%%%%%%%%%%%%

\maketitle
\begin{abstract}
The cross-media retrieval problem has received much attention in recent years due to the rapid increasing of multimedia data on the Internet. A new approach to the problem has been raised which intends to match features of different modalities directly. In this research, there are two critical issues: how to get rid of the heterogeneity between different modalities and how to match the cross-modal features of different dimensions. Recently metric learning methods show a good capability in learning a distance metric to explore the relationship between data points. However, the traditional metric learning algorithms only focus on single-modal features, which suffer difficulties in addressing the cross-modal features of different dimensions. In this paper, we propose a cross-modal similarity learning algorithm for the cross-modal feature matching. The proposed method takes a bilinear formulation, and with the nuclear-norm penalization, it achieves low-rank representation. Accordingly, the accelerated proximal gradient algorithm is successfully imported to find the optimal solution with a fast convergence rate $O(1/t^2)$. Experiments on three well known image-text cross-media retrieval databases show that the proposed method achieves the best performance compared to the state-of-the-art algorithms.
\end{abstract}

% A category with the (minimum) three required fields
%\category{H.3}{Information Storage and Retrieval}{Information
%Search and Retrieval}
%\category{H.3.3}{Information Systems}{Information Storage and Retrieval}[Information
%Search and Retrieval]
%A category including the fourth, optional field follows...
%\category{D.2.8}{Software Engineering}{Metrics}[complexity measures, performance measures]

%\terms{Information Retrieval}

\keywords{Multimedia Retrieval; Cross-Modality; Similarity Learning; Nuclear Norm; Accelerated Proximal Gradient.}

\section{Introduction}

A large amount of multimedia data has been released on the Internet during the past decade, and it is still rapidly increasing. From this big amount of media data, there exist some popular but different modalities, such as images, audios and documents. As a result, the requirement of cross-modal retrieval becomes a new problem to both researchers and the industry. Examples include the heterogenous face recognition (i.e. sketches and photos), cross-lingual retrieval, and the cross-media retrieval. Take the image-document cross retrieval as an example, given an image, the task is to find the documents that best describe it, or on the contrary, given several words, the task is to find the most related images.

Recently, Rasiwasia et al. \cite{Rasiwasia-mm-10} proposed a new approach to the multimedia information retrieval problem. In the research, the authors want to match features of different modalities directly, although the cross-modal data is in rather different feature spaces. The key problem to this research is to best reduce the heterogeneity among the different modal features, or find a common latent space in which the different modal features can be matched to each other.

%Classical algorithms have also been applied to solve the problem, such as the Canonical Correlation Analysis (CCA) \cite{Hardoon-neco-04,Rasiwasia-mm-10} and the Partial Least Squares (PLS) \cite{Rosipal-slsfs-06,Sharma-cvpr-12}. Specifically, the CCA aims at learning a latent space where the correlations between the projected features of the two modalities have been mutually maximized. The algorithm and its extensions have been widely used in multimedia field \cite{Hardoon-neco-04,hwang-bmvc-10,Rasiwasia-mm-10,Li-iccv-11}. Similar to the CCA, the PLS also learns a latent space, but with different formulations to extract latent vectors \cite{Rosipal-slsfs-06}.

Classical algorithms have been applied to solve the problem, such as the Canonical Correlation Analysis (CCA) \cite{Hardoon-neco-04,Rasiwasia-mm-10} and the Partial Least Squares (PLS) \cite{Rosipal-slsfs-06,Sharma-cvpr-12}. Specifically, the CCA aims at learning a latent space where the correlations between the projected features of the two modalities have been mutually maximized. Similar to the CCA, the PLS also learns a latent space, but with different formulations to extract latent vectors \cite{Rosipal-slsfs-06}. In the work of \cite{Rasiwasia-mm-10,jose-pami-2014}, Rasiwasia et al. proposed a semantic correlation matching (SCM) algorithm to deal with the problem. In the algorithm, a semantic level matching is suggested to combine with the linear correlations among the features. They also pointed out that, the correlation matching (such as CCA) is not enough for cross-modal matching, and the semantic level matching brings much benefit in working out the problem. This is consistent with that the class label information is very helpful to reduce the semantic gap, which is well known in the image retrieval field denoting the gap between the high level documents and low level images. Beyond these methods, there are some other algorithms to deal with the problem, such as the generalized multiview analysis (GMA) \cite{Sharma-cvpr-12}, the weakly paired maximum covariance analysis (WMCA) \cite{Lampert-eccv-10}, the deep neural network \cite{Ngiam-icml-11,Srivastava-nips-12}, etc. \cite{Bronstein-cvpr-10,pan-sigir-14,jia-iccv-11,zhuang-wsdm-11,zhuang-aaai-13,funZhu-cikm-14,gong-ijcv-14}.

%, domain transfer \cite{Duan-ICML-2012,kulis-cvpr-11,socher-nips-13}

Recently, metric learning algorithms have been popularly studied. The target of metric learning is to find a distance metric by using the similar constraints and the dissimilar constraints \cite{xing-nips-2002}. Following the work of Xing et al. \cite{xing-nips-2002}, there are some popular algorithms like the Large Margin Nearest Neighbor (LMNN) \cite{weinberger-nips-06}, the Information-Theoretic Metric Learning (ITML) \cite{davis-icml-07}, the Relevance Component Analysis (RCA) \cite{Bar-icml-03}, and the Discriminative Component Analysis (DCA) which is based on the RCA \cite{hoi-cvpr-06}. Specifically, the LMNN was proposed to use a large margin setting to improve the k-NN classification. The ITML algorithm proposed by Davis et al. aims at minimizing the differential relative entropy between two multivariate Gaussians. However, these traditional metric learning algorithms are designed for a single modality. They suffer the difficulty of learning the similarity/distance metrics between different modalities, especially with totaly different features.

Inspired by metric learning, we propose a novel logistic loss based cross-modality similarity function learning algorithm for the cross-media retrieval. The similarity formulation is in a bilinear form of two modalities, and a fast optimization algorithm is imported to find the optimal solution with a convergence rate of $O(1/ t^2)$, where $t$ is the number of iterations. Besides, the nuclear-norm regularization is imported in the formulation to explore the structures of the different modalities for robust learning. The proposed algorithm is applied to multimedia information retrieval on three popularly used multimedia databases for experiments, namely the Pascal VOC2007 database \cite{PASCAL-VOC2007,Sharma-cvpr-12}, the NUS-WIDE database \cite{nuswide-09}, and the Wikipedia database \cite{Rasiwasia-mm-10}. Experimental results show that the proposed algorithm is effective for cross-media retrieval. Especially, on the Wikipedia database the proposed algorithm outperforms the second best one for more than 6\%.

The rest of the paper is organized as following. In Section 2, a concise review is given for the related works in the cross-modal information retrieval field and the metric learning field. Then, in Section 3, after a brief introduction of the Metric Learning problem, the formulation of the proposed algorithm and how to find the optimal solution are introduced in detail. The experiments are described in Section 4 to validate the effectiveness of the proposed algorithm. Finally, Section 5 gives the conclusion of the paper.

\section{Related Work}

Information Retrieval (IR) is an important problem in the multimedia field. However, many traditional methods to the problem belong to the unimodal algorithms, such as the document retrieval and the content based image retrieval. Due to the rapid development of Internet applications, the cross-modality information retrieval becomes a common scene. To retrieve an image, the query terms may be of various different modalities, such as paragraphs, sketches and audios. Thus, cross-modal matching and learning algorithms which can match different modality data directly become a new research interest in the IR field.

Among these cross-modal matching algorithms, the CCA algorithm is the most widely used method in the multimedia field \cite{Hardoon-neco-04,hwang-bmvc-10,Rasiwasia-mm-10,Li-iccv-11}. The target of CCA is to learn a latent space by maximizing the correlating relationships between two modality features. Thus the different modal features can be projected to the latent space for similarity computation. The algorithm is also used as the correlation matching (CM) method by Rasiwasia et al. \cite{Rasiwasia-mm-10}. Beyond CCA, the PLS algorithm is another classical method for cross-modal data \cite{Rosipal-slsfs-06,Sharma-cvpr-11,Sharma-cvpr-12}. The core idea of it is very similar with that of CCA, which is to extract the latent vectors with maximal correlations.

In \cite{Rasiwasia-mm-10} where the cross-modal IR was suggested, Rasiwasia et al. proposed a supervised algorithm for the image-text cross-modal retrieval problem, namely the SCM algorithm. The SCM is one of the most famous and the current state-of-the-art algorithm. To reduce the semantic gap between images and documents, the sematic level matching is developed based on the learned maximal correlation latent space by CCA. Thus, the algorithm can be separated into two steps. The correlational matching between different modalities by CCA is done in the first step. Then, based on it a semantic space is learned in the second step. As indicated in \cite{Rasiwasia-mm-10,jose-pami-2014}, the class information is an important information to reduce the semantic gap. Thus, in order to use the class labels, Sharma et al. proposed a GMA algorithm to learn a discriminative latent space for cross-modal data and treat it as an eigenvalue problem. The algorithm shows great performance to the pose and lighting invariant face recognition and cross-modal retrieval problems.

In fact, some methods targeting to the heterogenous face recognition problem are available to deal with the cross-modality IR problem, such as the Multiview Discriminant Analysis (MvDA) method \cite{Kan-eccv-12}. The MvDA algorithm aims at learning a common space where the between-class variations from both inter-view and intra-view are maximized, and the within-class variations from both inter-view and intra-view are minimized. In the transfer learning field, some algorithms are also related \cite{Duan-ICML-2012,kulis-cvpr-11,Lampert-eccv-10,socher-nips-13}. Such as in \cite{Lampert-eccv-10}, Lampert and Kr\"{o}mer proposed a weakly-paired maximum covariance analysis method to deal with the not fully paired (not one-by-one paired) training data. Besides, Wang et al. \cite{wang-iccv-13} proposed an iterative algorithm based on sparsity to learn the coupled feature spaces for the different modalities. The work in \cite{funZhu-cikm-14} also proposed a greedy dictionary construction approach to select dictionary atoms for constructing a modality-adaptive dictionary pair. In the deep learning field, the works \cite{Ngiam-icml-11} and \cite{Srivastava-nips-12} both used the restricted boltzmann machine for the cross-modal feature learning.

In the literature of the metric learning field, Chechik et al. \cite{gal-nips-09} imported an online similarity function learning for large-scale images. But the algorithm is designed for single-modality in a triplet ranking formulation. Kulis et al. also proposed to learn an asymmetric transformation matrix for domain adaption \cite{kulis-cvpr-11}. Besides, a metric learning algorithm for different modalities was also realized by Wu et al. \cite{wu-report-10}. The objective function in their work learns the projections for the different modalities, respectively, to best separate the similar points set and the dissimilar points set. However, the pair-wise information is ignored in the algorithm. In \cite{zhai-aaai-13} and \cite{mignon-accv-12}, the authors also proposed to learn two projections for each modality to minimize the distances of the two modalities in the target feature space. Zhai et al. also used the semantic information in the second step based on a unified k-NN graph. However, both of the algorithms are not convex, thus the optimal solution is not guaranteed. In contrast, the proposed formulation in this paper is a strict convex problem, and the optimal solution is achieved by the accelerated proximal gradient (APG) algorithm \cite{Nes-book-03,Tseng-SIAM-08,beck-siam-09}.

\section{Similarity Learning for Cross Modalities}

\subsection{Traditional Metric Learning}
%As we have said in the introduction, the Metric Learning was firstly proposed by Xing \textit{et al.} \cite{xing-nips-2002} to learn a distance metric between the point pairs according to the similar points.

In this section, we give a brief review of the traditional metric learning problem, which was firstly studied by Xing et al. \cite{xing-nips-2002} to learn a distance metric according to the similar points and dissimilar points.

Suppose that there is a set of $m$ data points $\{ \mathbf{x}_i \in \mathds{R}^{d}\}_{i=1}^{m}$, where each data point is with a $d$ dimensional feature. Besides, we are also given the binary similarity information indicating whether two data points are similar or not. Specifically, there are two sets of paired points, which can be described as follows:
\begin{equation}
\begin{cases}
\mathcal{S} = \{(\mathbf{x}_i, \mathbf{x}_j) \mid \text{$\mathbf{x}_i$ and $\mathbf{x}_j$ are in the same class}\}, \\
\mathcal{D} = \{(\mathbf{x}_i, \mathbf{x}_j) \mid \text{$\mathbf{x}_i$ and $\mathbf{x}_j$ are not in the same class}\},
\end{cases}
\end{equation}
where $\mathcal{S}$ is called the must-link set containing the similar pairs and the $\mathcal{D}$ is called the cannot-link set containing the dissimilar pairs.

Given the two sets, the task of metric learning is to learn a distance metric in the following form,
\begin{equation}
d(\mathbf{x}_i, \mathbf{x}_j)_{\mathbf{M}} = \|\mathbf{x}_i - \mathbf{x}_j \|_{\mathbf{M}}  = \sqrt{(\mathbf{x}_i - \mathbf{x}_j)^{T}\mathbf{M}(\mathbf{x}_i - \mathbf{x}_j)},
\end{equation}
where $\mathbf{M}\in\mathds{R}^{d\times d}$ is a Mahalanobis distance metric that should be learned. It can be easily found that $\mathbf{M}$ is symmetric. Furthermore, to be a valid metric, $\mathbf{M}$ must satisfy the non-negativity and the triangle inequality. Thus $\mathbf{M}$ is required to be positive semi-definite, namely $\mathbf{M}\succeq0$.

Inspired by the metric learning formulation, we want to learn a similarity function to evaluate two modality features which are in different dimensions, for example the $\langle image, text\rangle$ pairs. In this way, the different modality features can be matched directly with the learned similarity function. The proposed formulation is introduced as follows.

\subsection{Problem Definition and Formulation}
Suppose that there is a multimedia database $\mathcal{A} = \{(\mathbf{x}_1, l_1^{x}),$ $(\mathbf{x}_2, l_2^{x}),\ldots,(\mathbf{x}_m, l_m^{x}), (\mathbf{z}_1, l_1^{z}), \ldots, (\mathbf{z}_n, l_n^{z})\}$ constructed from two-modality data sets $\mathcal{X}$ and $\mathcal{Z}$. Sample $\mathbf{x}_i \in \mathds{R}^{d_1}$ is from the modal $\mathcal{X}$, such as the image, document, or audio, and $\mathbf{z}_i \in \mathds{R}^{d_2}$ belongs to the modality $\mathcal{Z}$ which is different from the modality $\mathcal{X}$, where $d_1$ and $d_2$ are dimensions of the two modality features, respectively. The $l_i^{x}$ and $l_j^{z}$ are the class labels of the samples $\mathbf{x}_i$ and $\mathbf{z}_j$ respectively. The connection between the two modalities is that they share $c$ classes, for example, dog, shoes and buildings. The cross matching problem is how to match the cross-modal samples $\mathbf{x}_i$ and $\mathbf{z}_j$ directly.% For example, given an image, find the documents to best describe the image, or given several words, find the most related images.

Similar with the metric learning, we also define two pairwise sets on the cross-modal samples,
\begin{equation}
\begin{cases}
\mathcal{S} = \{(\mathbf{x}_i, \mathbf{z}_j) \mid l_i^{x}=l_j^{z}\}, \\
\mathcal{D} = \{(\mathbf{x}_i, \mathbf{z}_j) \mid l_i^{x}\neq l_j^{z}\},
\end{cases}
\end{equation}
where $\mathcal{S}$ is the must-link set with similar pairs from the two different modalities, and $\mathcal{D}$ is the cannot-link set with dissimilar pairs from the two modalities.

Based on these definitions, we want to learn a similarity function which can measure the similarity between any two different modal features. The formulation of the proposed algorithm is
\begin{equation}
\label{eqn:logis-rank}
\min_{\mathbf{M}}\sum_{i,j=1}^{m,n} w_{ij}\log(1+\exp(-y_{ij} S_{\mathbf{M}}(\mathbf{x}_i, \mathbf{z}_j)) ) + \lambda\|\mathbf{M}\|_{*},
\end{equation}
where $S_{\mathbf{M}}(\mathbf{x}_i, \mathbf{z}_j)$ is the similarity function parameterized by a matrix $\mathbf{M}\in \mathds{R}^{d_1\times d_2}$, which is to be learned in a bilinear form,
\begin{equation}
S_{\mathbf{M}}(\mathbf{x}_i, \mathbf{z}_j) = \mathbf{x}_i^{T} \mathbf{M} \mathbf{z}_j.
\end{equation}

The formulation in Eqn.(\ref{eqn:logis-rank}) contains a loss function item and a regularization item. The loss function is based on the logistic sigmoid function \cite{guillaumin-iccv-09}, $\delta(x) = 1/(1+\exp(-x))$, which is often integrated in $-\log(\delta(x))$ to approximate the hinge loss function to avoid the discontinuity of the gradient. Here,  $y_{ij} \in \{+1, -1\}$ is a sign indicating whether the pair is similar (positive) or dissimilar (negative). Specifically,
\begin{equation}
y_{ij} =
\begin{cases}
+1 \hspace{5mm} (\mathbf{x}_i, \mathbf{z}_j) \in \mathcal{S}, \\
-1 \hspace{5mm} (\mathbf{x}_i, \mathbf{z}_j) \in \mathcal{D}.
\end{cases}
\label{eqn:LRBS-yij}
\end{equation}
For $w_{ij}$ in the formulation, it stands for the weight of the $(\mathbf{x}_i, \mathbf{z}_j)$ pair, which is used to deal with unbalanced positive and negative samples. In our experiments, we set the weights for positive (negative) pairs to be the reciprocal of the number of positive (negative) pairs.

As for the regularization term in Eqn.(\ref{eqn:logis-rank}), the nuclear norm $\|\cdot\|_{*}$ is used, which is defined as the sum of the singular values of a matrix. The functions of the nuclear norm in the proposed formulation lie in two folds. On one fold, the constraint can be treated as a scale regularization of $\mathbf{M}$. It makes the solution to Eqn.(\ref{eqn:logis-rank}) in a constrained domain. On the other fold, the nuclear norm can make the learned metric matrix $\mathbf{M}$ with a low rank, which is a desirable property and has been widely used in the machine learning field. In the problem Eqn.(\ref{eqn:logis-rank}), the nuclear norm may help to discover the connections between the two modalities and result in a robust learning.

\subsection{Optimization}
In this subsection, the accelerated proximal gradient algorithm (APG) \cite{Nes-book-03,Tseng-SIAM-08} is utilized to find the optimal solution of Eqn.(\ref{eqn:logis-rank}). The APG algorithm is a kind of first order gradient descent method \cite{beck-siam-09,Ji-ICML-2009,Nes-book-03,Tseng-SIAM-08}, which has received popular attentions in recent years due to its fast convergence rate of $O(1/t^2)$ when dealing with the problem of convex and smooth loss function regularized by non-smooth constraints.

For convenience, the formulation in Eqn.(\ref{eqn:logis-rank}) can be rewritten as
\begin{equation}
\label{eqn:logis-rank2}
\min_{\mathbf{M}} f(\mathbf{M}) = l(\mathbf{M}) + \lambda\|\mathbf{M}\|_{*},
\end{equation}
where
\begin{equation}
l(\mathbf{M}) =  \sum_{i,j=1}^{m,n} w_{ij}\log(1+\exp(-y_{ij} \mathbf{x}_i^{T}\mathbf{M} \mathbf{z}_j) ).
\end{equation}
Note that $l(\mathbf{M})$ is a smooth and convex function, and the objective function is convex. Accordingly, the APG algorithm can be used to solve the proposed formulation.

To find the optimal minimizer, we first construct a proximal operator of the Eqn. (\ref{eqn:logis-rank2}), which is
\begin{equation}
\centering
\begin{split}
\min_{\mathbf{M}} g(\mathbf{M}) & = l(\mathbf{Q}_t) + \langle  l'(\mathbf{Q}_t), \mathbf{M} -\mathbf{Q}_t \rangle \\ & + \frac{1}{2\eta_{t}} \|\mathbf{M} - \mathbf{Q}_t\|_2^2 + \lambda\|\mathbf{M}\|_{*},
\end{split}
\label{eqn:operator_g}
\end{equation}
where $\{\mathbf{Q}_t\}$ is a sequence of the search points, $l'(\mathbf{Q}_t)$ is the gradient of $l(\cdot)$ at the $\mathbf{Q}_t$ point, and $\eta_{t}$ is the update step size at the $t$-th iteration. The $l'(\mathbf{Q}_{t})$ is
\begin{equation}
\label{eqn:loss-grad}
\centering
\frac{\partial l(\mathbf{M})}{\partial\mathbf{M}}|_{\mathbf{Q}_t} = - \sum_{i,j=1}^{m,n} \frac{ w_{ij}y_{ij} }{1+\exp(y_{ij} \mathbf{x}_i^{T}\mathbf{Q}_{t} \mathbf{z}_j) } \mathbf{x}_i\mathbf{z}_j^{T},
\end{equation}

By defining a matrix $\mathbf{T}$ with each entry $T_{ij}$ as
\begin{equation}
\centering
T_{ij} =  \frac{ w_{ij}y_{ij} }{1+\exp(y_{ij} \mathbf{x}_i^{T}\mathbf{Q}_t \mathbf{z}_j) },
\end{equation}
we can conveniently compute $\mathbf{T}$ as
\begin{equation}
\centering
\mathbf{T} = \mathbf{W}\odot\mathbf{Y}\odot \frac{1}{1+\exp(\mathbf{Y}\odot\mathbf{X}^{T}\mathbf{Q}_t \mathbf{Z}) },
\end{equation}
where $\odot$ is the element-wise product of two matrices. $\mathbf{W}\in \mathds{R}^{m\times n}$ is the weight matrix with $w_{ij}$ for each pair of the cross-modal samples, and $\mathbf{Y}\in \mathds{R}^{m\times n}$ is the sign matrix with $y_{ij}$ defined in Eqn.(\ref{eqn:LRBS-yij}). The $\mathbf{X}\in \mathds{R}^{d_1\times m}$ is the data matrix containing the $m$ samples of the modality $\mathcal{X}$, and $\mathbf{Z}\in \mathds{R}^{d_2\times n}$ consists of the $n$ samples of the modality $\mathcal{Z}$.

With these definitions, Eqn.(\ref{eqn:loss-grad}) can be simplified as
\begin{equation}
\centering
\begin{split}
l'(\mathbf{Q}_t) = \frac{\partial l(\mathbf{M})}{\partial\mathbf{M}}|_{\mathbf{Q}_t} & = - \sum_{i=1}^{m}\sum_{j=1}^{n} T_{ij} \mathbf{x}_i\mathbf{z}_j^{T}\\
 & = - \mathbf{X} \mathbf{T} \mathbf{Z}^{T}
\end{split}
\end{equation}

By removing the constant term $l(\mathbf{Q}_t)$ and adding another constant term $\eta_{t}\|l'(\mathbf{Q}_t)\|_2^2/2$ with respect to $\mathbf{M}$, the proximal operator of Eqn. (\ref{eqn:operator_g}) is equal to
\begin{equation}
\label{eqn:square-rank}
\centering
\min_\mathbf{M} \frac{1}{2} \|\mathbf{M} - \mathbf{\tilde{M}}_t\|_{F}^{2} + \lambda\eta_{t} \|\mathbf{M}\|_{*},
\end{equation}
where
\begin{equation}
\centering
\mathbf{\tilde{M}}_t =\mathbf{Q}_t -\eta_{t}l'(\mathbf{Q}_t).
\end{equation}
According to \cite{cai-siam-10}, the minimization of the objective function Eqn.(\ref{eqn:square-rank}) can be solved by a soft-thresholding technique. Firstly, the singular value decomposition (SVD) is used to find the eigenvalues for the matrix $\mathbf{\tilde{M}}$. Then, a soft-thresholding is applied on the computed eigenvalues. In summary,

%---------------------------------------
\newtheorem{theorem}{Theorem}
\begin{theorem}
Assume matrix $\mathbf{L} \in \mathds{R}^{m\times n}$ and the SVD decomposition of it is $\mathbf{L} =\mathbf{U}\mathbf{\Sigma}\mathbf{V}^{T}$, where $\mathbf{U}\in\mathds{R}^{d_1\times r}$ and $\mathbf{V}\in\mathds{R}^{d_2\times r}$ are constructed by orthogonal vectors, and $\mathbf{\Sigma}$ is a diagonal matrix with diagonal values $\Sigma_{ii}\geq0$ and $rank(\mathbf{\Sigma})=r$. Then the following objective function \cite{cai-siam-10}
\begin{equation}
\centering
\label{eqn:square-rank2}
C_{\gamma}(\mathbf{M}) = \min_\mathbf{M} \frac{1}{2} \|\mathbf{M} -\mathbf{L}\|_{F}^{2} + \gamma \|\mathbf{M}\|_{*}
\end{equation}
has a closed-form solution $C_{\gamma}(\mathbf{M})=\mathbf{U}\mathbf{\Sigma}_{\gamma}\mathbf{V}^{T}$, where $(\Sigma_{\gamma})_{ii}= \max\{0, \Sigma_{ii}-\gamma\}$ is the soft-thresholding result of the matrix $\mathbf{\Sigma}$ by the regularization parameter $\gamma$ .
\end{theorem}
%

%%---------------------------------------

%

The proof can be found in the Appendix. Accordingly, we get the $\mathbf{M}_{t+1}=\mathbf{U}\mathbf{\Sigma}_{\lambda\eta_t}\mathbf{V}^{T}$. Then, the APG algorithm accelerates the proximal gradient descent by updating $\mathbf{Q}_{t+1}$ with
\begin{equation}
\centering
\label{eqn:search-upt}
\mathbf{Q}_{t+1} = \mathbf{M}_{t+1} + \frac{\alpha_t -1}{\alpha_{t+1}} (\mathbf{M}_{t+1} - \mathbf{M}_{t} )
\end{equation}
where $\alpha_{t+1} = (1+\sqrt{1+4\alpha_{t}^2})/2$ and $\alpha_1=1$. In fact, the $\alpha_t$ can be updated in other ways if the certain conditions are satisfied \cite{Nes-book-03,Tseng-SIAM-08}. As for the step size $\eta_t$, it can be estimated in the algorithm by comparing the objective function and its proximal operator \cite{beck-siam-09}, whose derivation is omitted here due to the paper length.

It turns out that the APG algorithm updates $\mathbf{M}_t$ and $\mathbf{Q}_t$ iteratively to find the optimal solution, and the searching point $\mathbf{Q}_t$ is a linear combination of the latest two solutions of $\mathbf{M}_t$ and $\mathbf{M}_{t-1}$. The steps of the proposed algorithm is summarized in Algorithm \ref{alg:apg1}.

\begin{algorithm}[t]
\label{alg:apg1}
\caption{Optimization for Eqn. (\ref{eqn:logis-rank}).}

\KwIn{$\mathbf{X}\in\mathds{R}^{d_1\times m}$, $\mathbf{Z}\in\mathds{R}^{d_2\times n}$, and $\mathbf{Y}\in\mathds{R}^{m\times n}$.}
\KwOut{ $\mathbf{M}\in \mathds{R}^{d_1\times d_2}$}
\textbf{Initialization} $\mathbf{Q}_t = \mathbf{0}, t=1$\;
\Repeat{convergence}
{
    Compute $\mathbf{\tilde{M}}_t$ as in Eqn. (12) (13) (15)\;
    Decompose $\mathbf{\tilde{M}}_t =\mathbf{U}\mathbf{\Sigma}\mathbf{V}^{T}$\;
    Update $\mathbf{M}_{t+1}=\mathbf{U}\mathbf{\Sigma}_{\lambda\eta_t}\mathbf{V}^{T}$\;
    Update $\mathbf{Q}_{t+1}$ as in Eqn. (\ref{eqn:search-upt}).
}
\end{algorithm}

\section{Experiment}
In this section, the proposed low rank bilinear similarity learning algorithm (denoted as LRBS) is compared with several popular and state-of-the-art algorithms by experiments on three famous image-text databases.

%\subsection{Compared Methods and Evaluation Metrics}
\textbf{Compared Methods: }To evaluate the performance of the proposed algorithm, several famous algorithms in the cross-media retrieval field were compared in the experiments, including the popular methods CCA~\cite{Hardoon-neco-04,Rasiwasia-mm-10} and PLS~\cite{Rosipal-slsfs-06,Sharma-cvpr-11}, and the state-of-the-art methods SCM \cite{Rasiwasia-mm-10}, the Microsoft algorithm (MsAlg) \cite{wu-report-10}, and the Generalized Multiview Analysis methods, including the GMLDA (Generalized Multiview Linear Discriminant Analysis) and the GMMFA (Generalized Multiview Marginal Fisher Analysis) \cite{Sharma-cvpr-12}. A brief introduction of these algorithms can be found in the related work section.

\textbf{Evaluation Metrics: }For the evaluation, the mean average precision (MAP) and the precision-recall curve metrics were used in the experiments. In fact, both of them are popularly used in evaluation of information retrieval systems. Given one query and its N retrieved documents in a rank, the average precision is computed by AveP = $\frac{1}{T}\sum_{r=1}^{N}P(r)rel(r)$, where $T$ is the number of the relevant documents in the retrieval, $P(r)$ is the precision of the top $r$ retrieved documents, and $rel(r)$ is a binary function denoting whether the $r$-th retrieved document is relevant to the query or not. Having the AveP calculated for each query, the MAP is computed as the average AveP score of all queries. Besides of the MAP and precision-recall curve, we also displayed the precision-scope curve for a better visualization, where the scope denotes the number of retrieved documents. Therefore, by the precision-scope curve it is more easily to see the precision of the top $r$ documents presented to the users.

\begin{table}[t]
\center
\caption{\small{MAP (\%) results on the Pascal VOC2007 database.}}
\begin{tabular}{ l|c|c|c}
\hline
                & \multicolumn{3}{c}{Without PCA}        \\  \hline
Methods         & Image query & Text query  & ~Average~  \\    \hline
CCA/SCM         &   --         &  --         & --         \\
PLS             & 47.55        & 45.69       & 46.62      \\
MsAlg           & 58.43        & 59.40       & 58.91      \\
GMLDA           & 39.95        & 35.95       & 37.95     \\
GMMFA           & 37.97        & 33.17       & 35.57     \\
LRBS             & 65.15        &\textbf{68.74} &\textbf{66.95} \\
\hline  \hline
                & \multicolumn{3}{c}{With PCA}        \\ \hline
Methods         & Image query & Text query  & ~Average~  \\    \hline
CCA             & 49.06        & 48.33       & 48.69      \\
PLS             & 47.55        & 45.68       & 46.62      \\
SCM             & 63.98        & 59.73       & 61.86    \\
MsAlg           & 58.58        & 58.84       & 58.71    \\
GMLDA         &\textbf{65.62}  & 66.32       & 65.97     \\
GMMFA          &{\color{blue}\textbf{65.48}} & 66.15       & 65.81     \\
LRBS             & 65.10        & {\color{blue}\textbf{68.69}} & {\color{blue}\textbf{66.90}}  \\
\hline
\end{tabular}
\label{tlb:voc-rst}
\end{table}
\begin{figure*}[t]
\center
\includegraphics[width=2.50in]{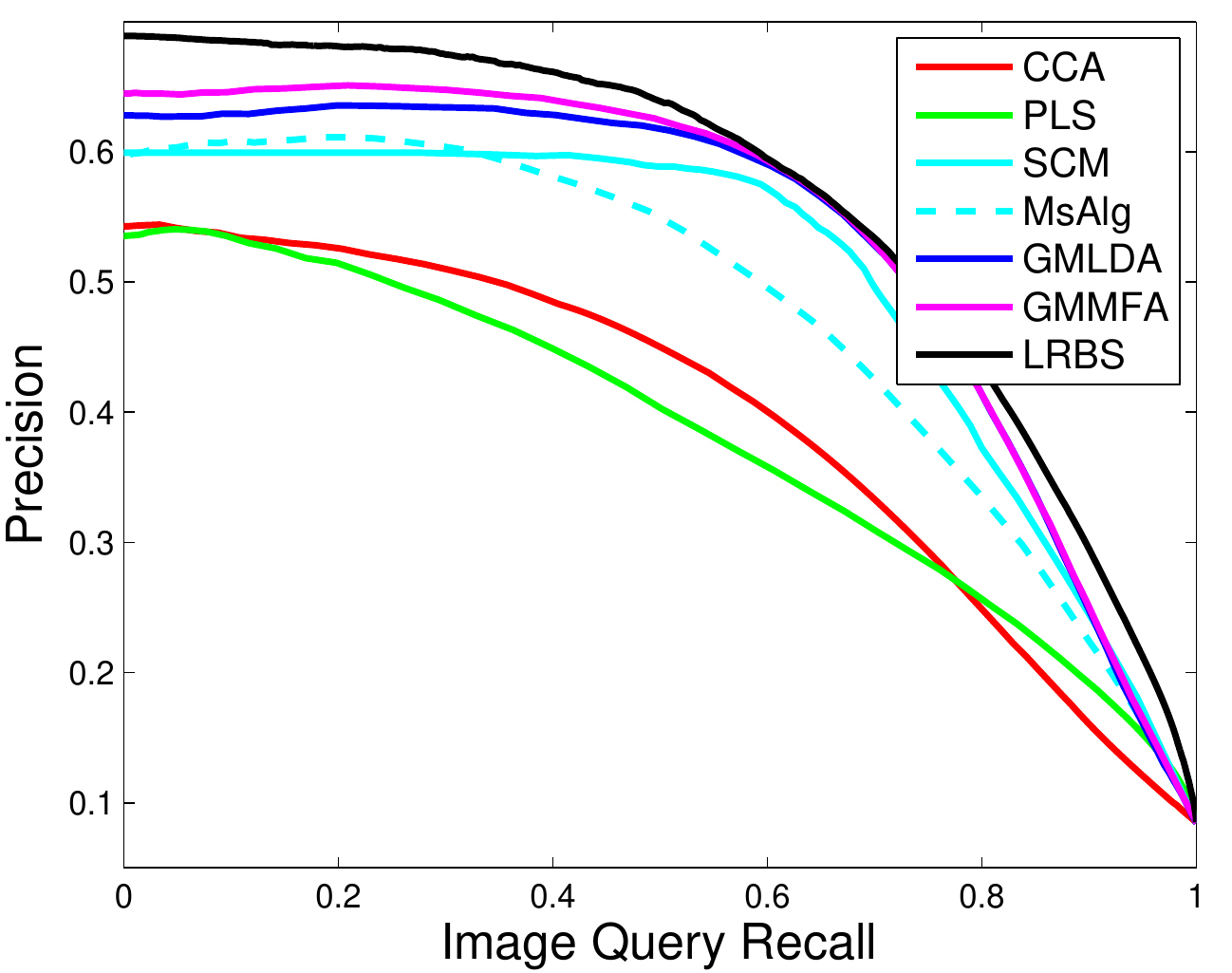}\hspace{12mm}
\includegraphics[width=2.50in]{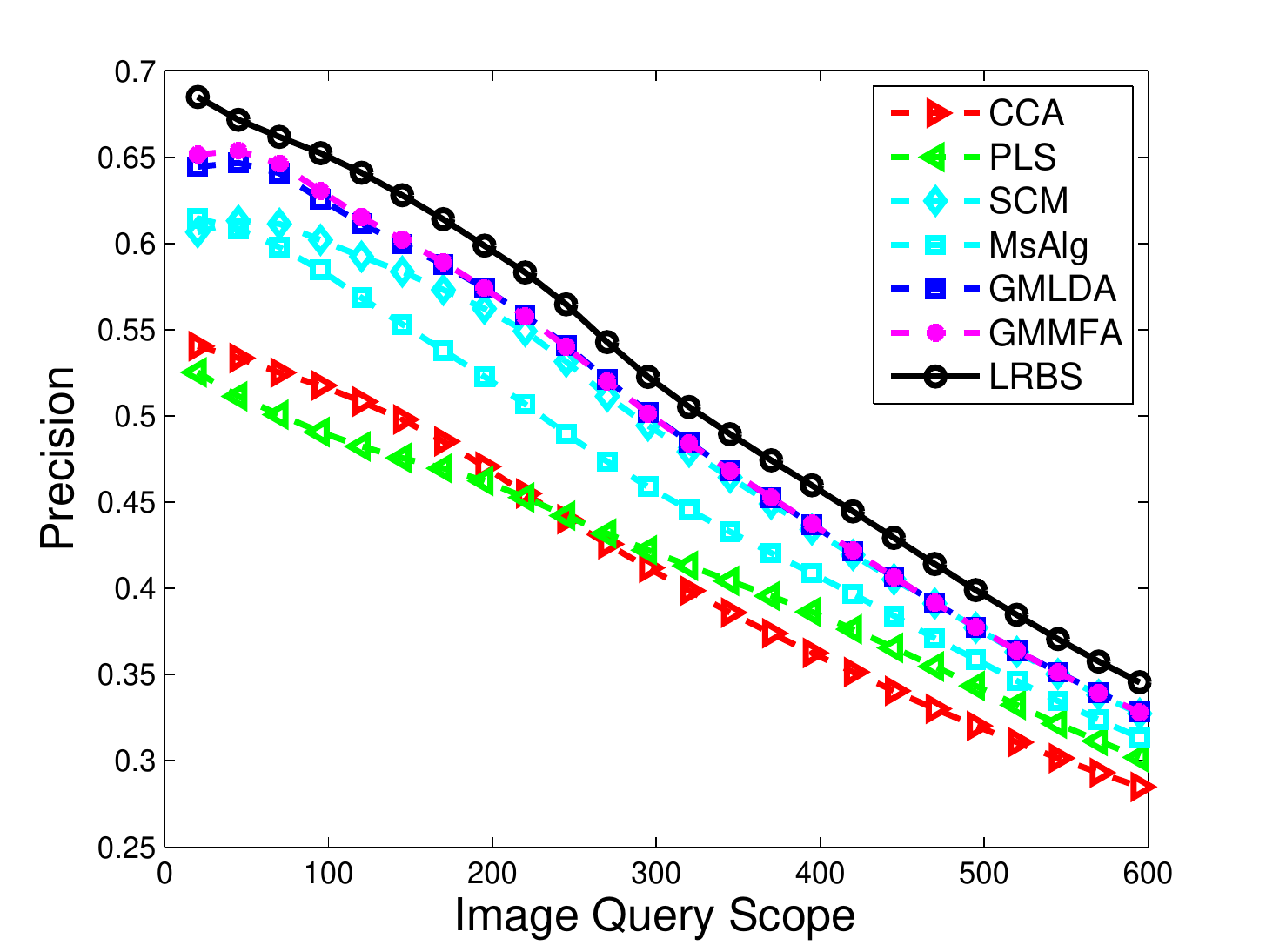} \\
(a) Image-to-Text \\
%\vspace{1ex}
\includegraphics[width=2.50in]{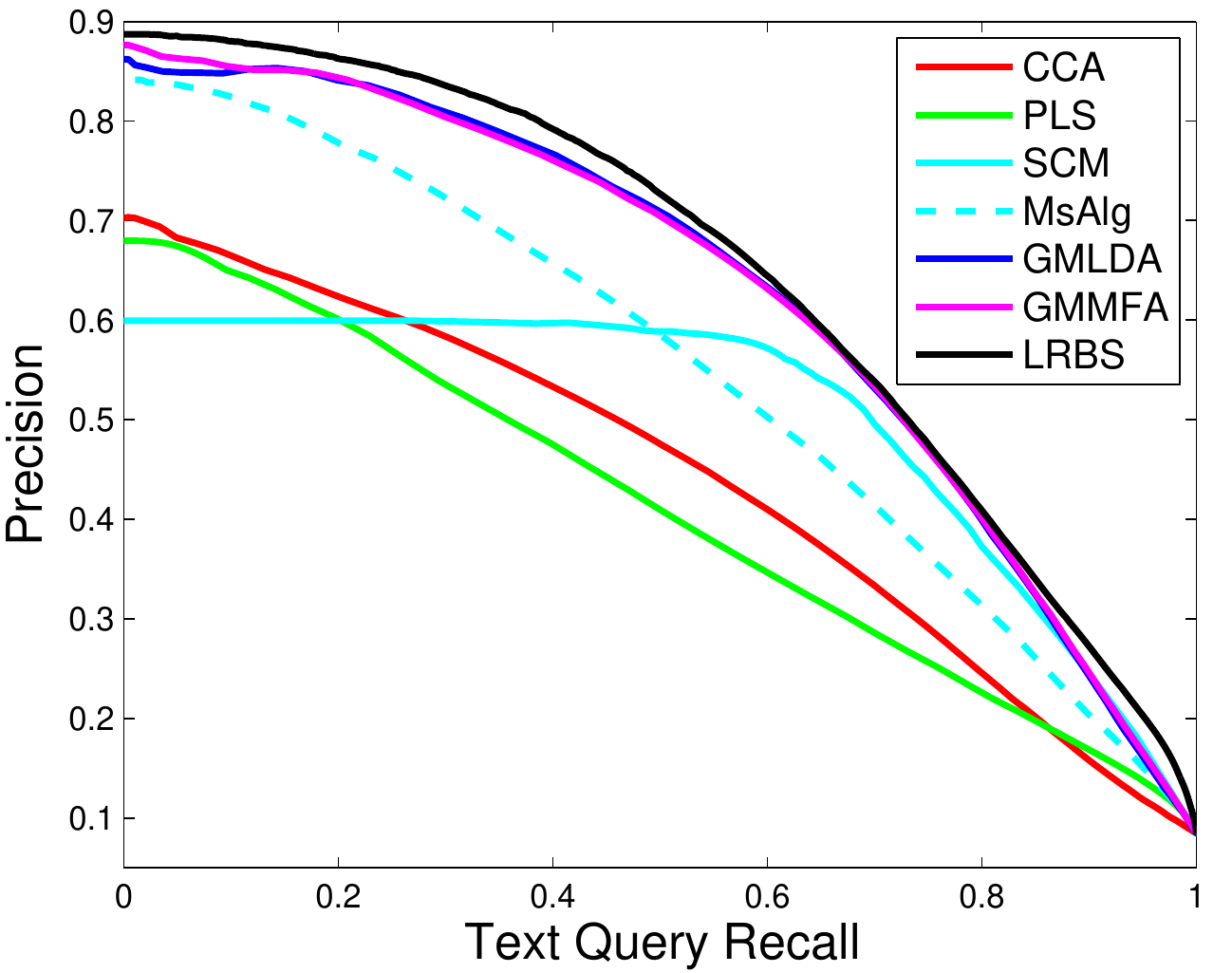} \hspace{12mm}
\includegraphics[width=2.50in]{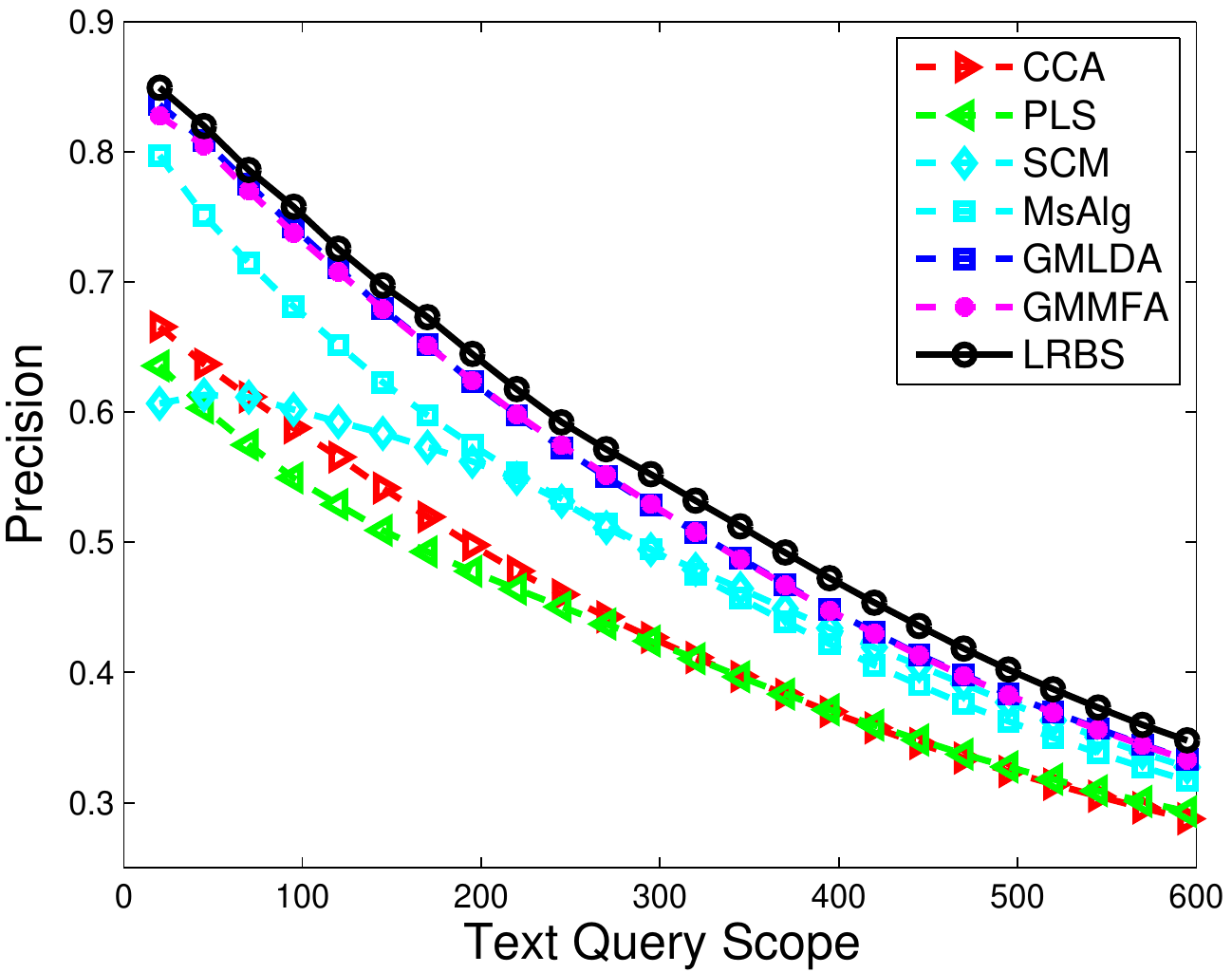}\\
(b) Text-to-Image
\caption{Precision-recall and precision-scope curves on the Pascal VOC2007 database for the Image-to-Text retrieval experiment and the Text-to-Image retrieval experiment.}
\label{fig:voc-rst}
\end{figure*}
% The 1st row shows the precision-recall curve, and the 2nd row shows the precision-scope curve. On the left of each row, it is the Image-to-Text retrieval results, and on the right is the Text-to-Image retrieval results
%
\begin{figure*}[!htb]
\center
\includegraphics[width=2.50in]{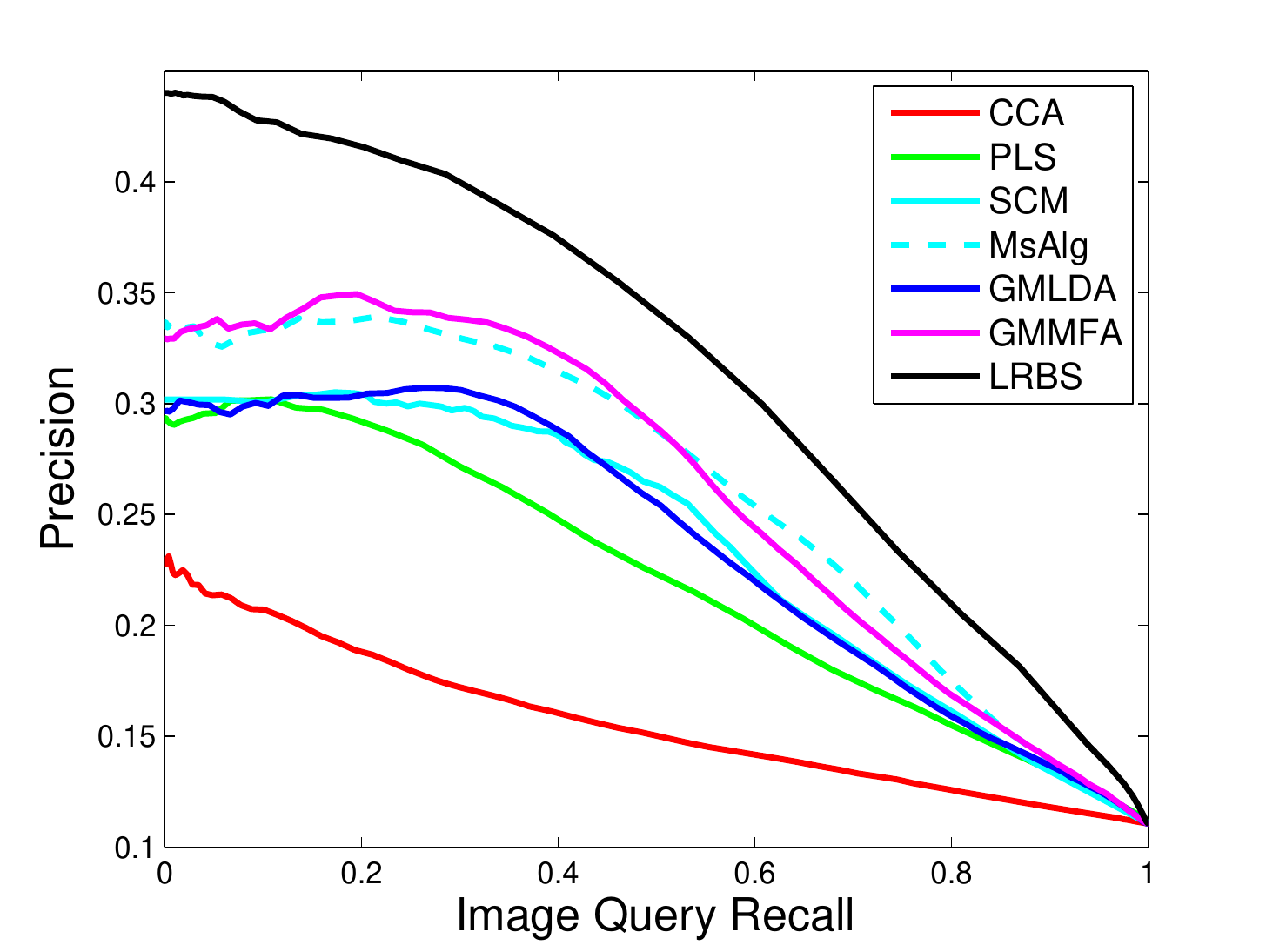}\hspace{12mm}
\includegraphics[width=2.50in]{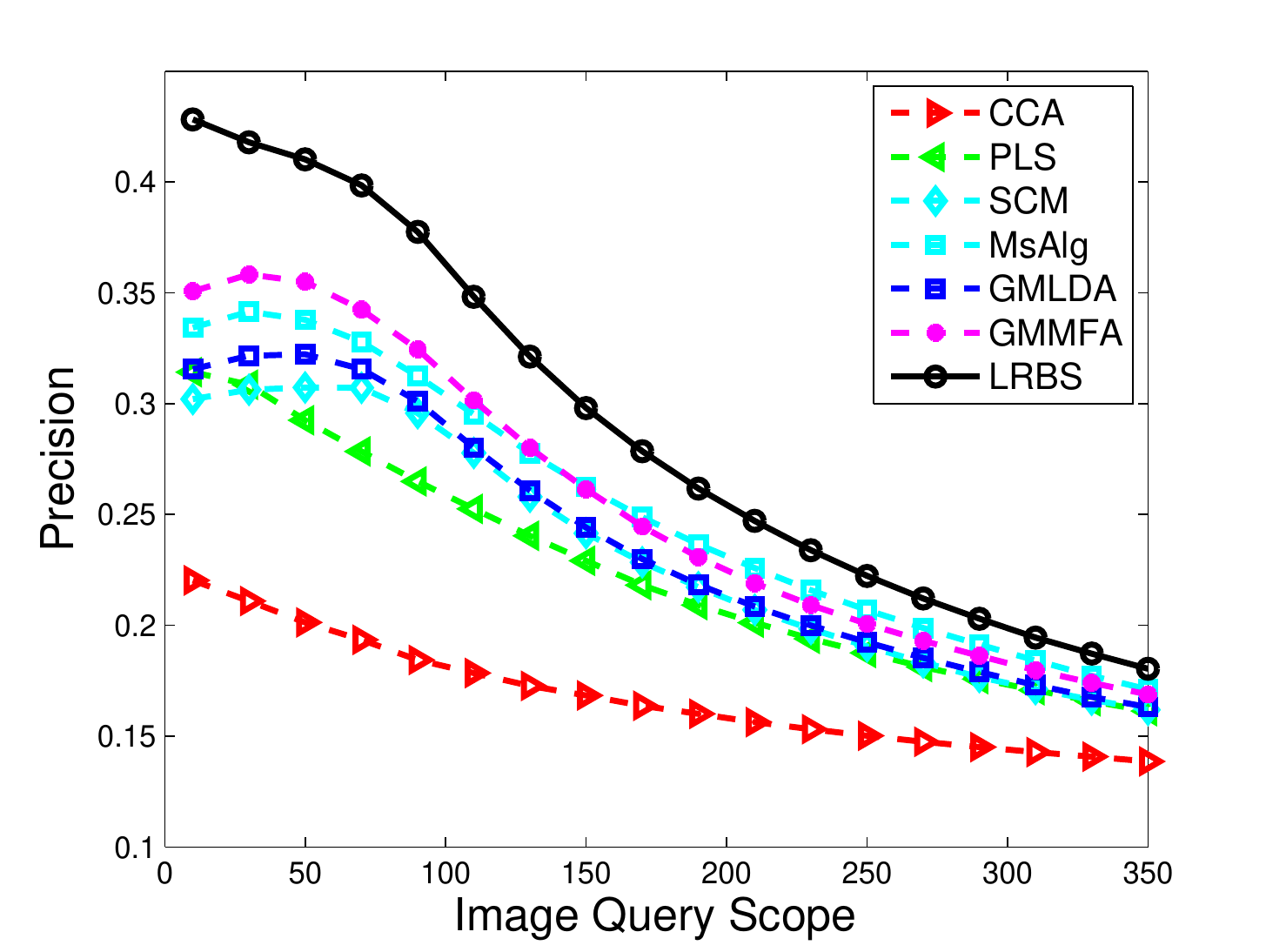} \\
(a) Image-to-Text \\
%\vspace{1ex}
\includegraphics[width=2.50in]{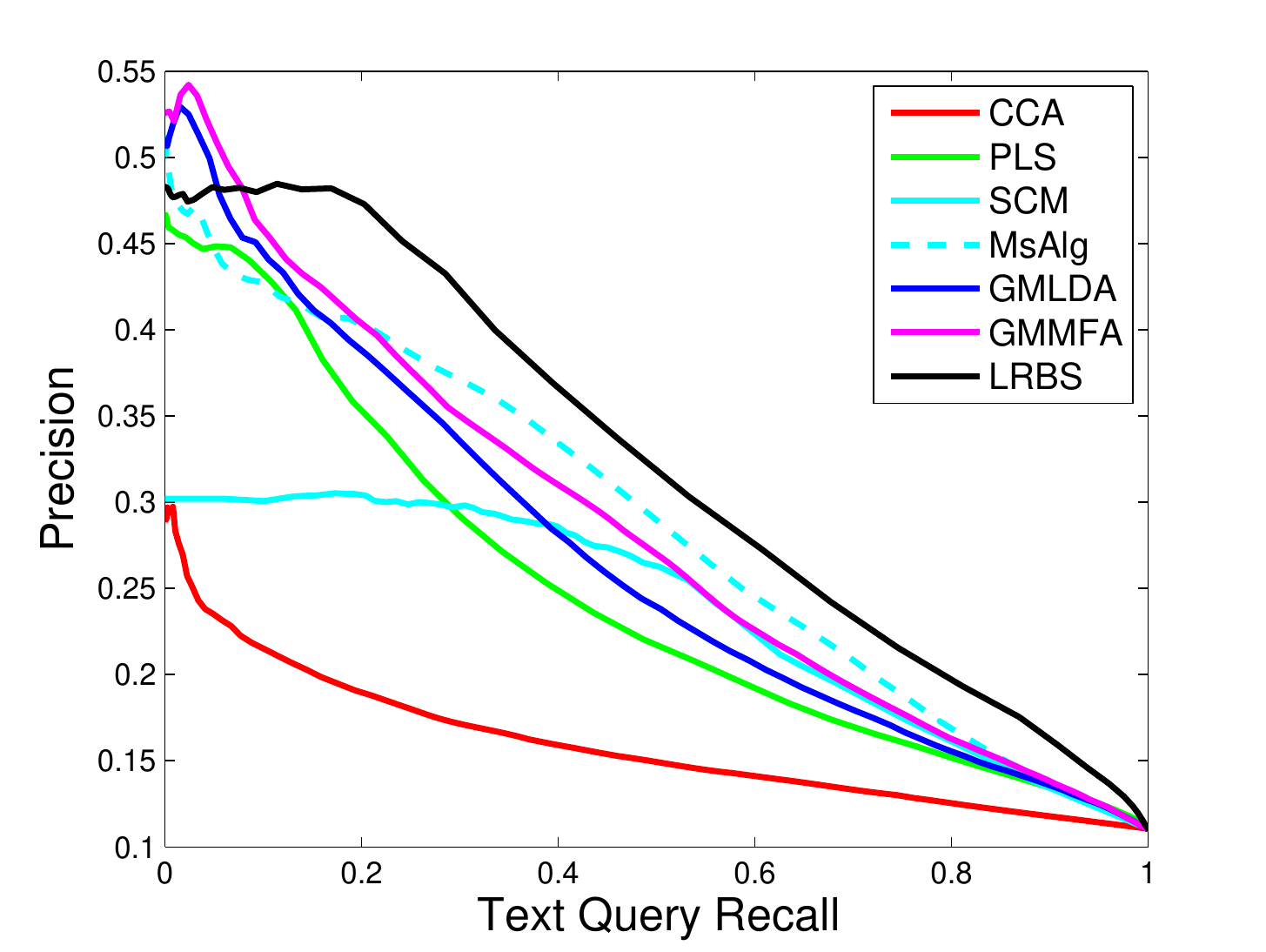} \hspace{12mm}
\includegraphics[width=2.50in]{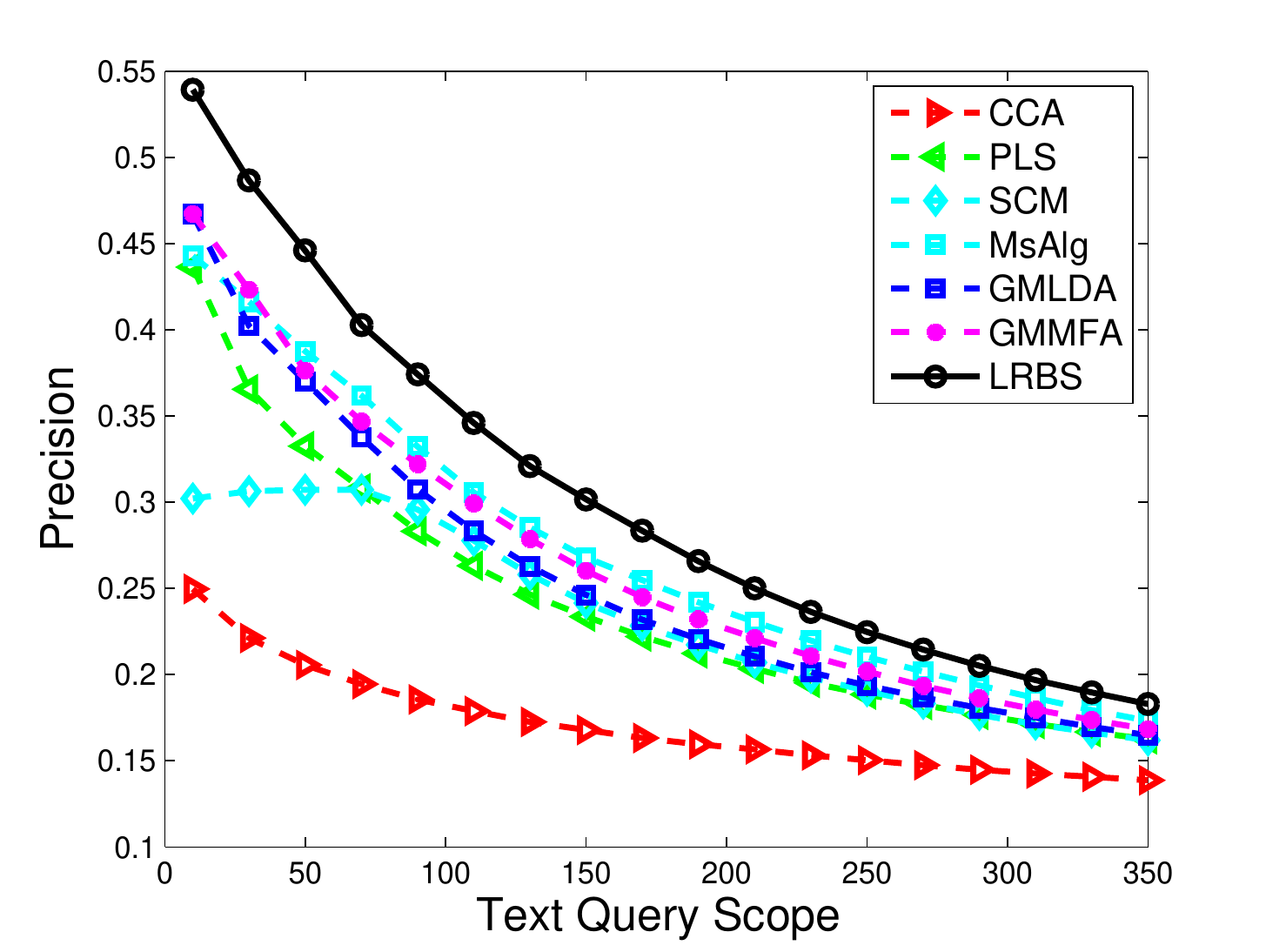}\\
(b) Text-to-Image
\caption{Precision-recall and precision-scope curves on the Wikipedia database for the Image-to-Text retrieval experiment and the Text-to-Image retrieval experiment.}
\label{fig:wiki-rst}
\end{figure*}

\begin{figure*}[!htb]
\center
\includegraphics[width=2.50in]{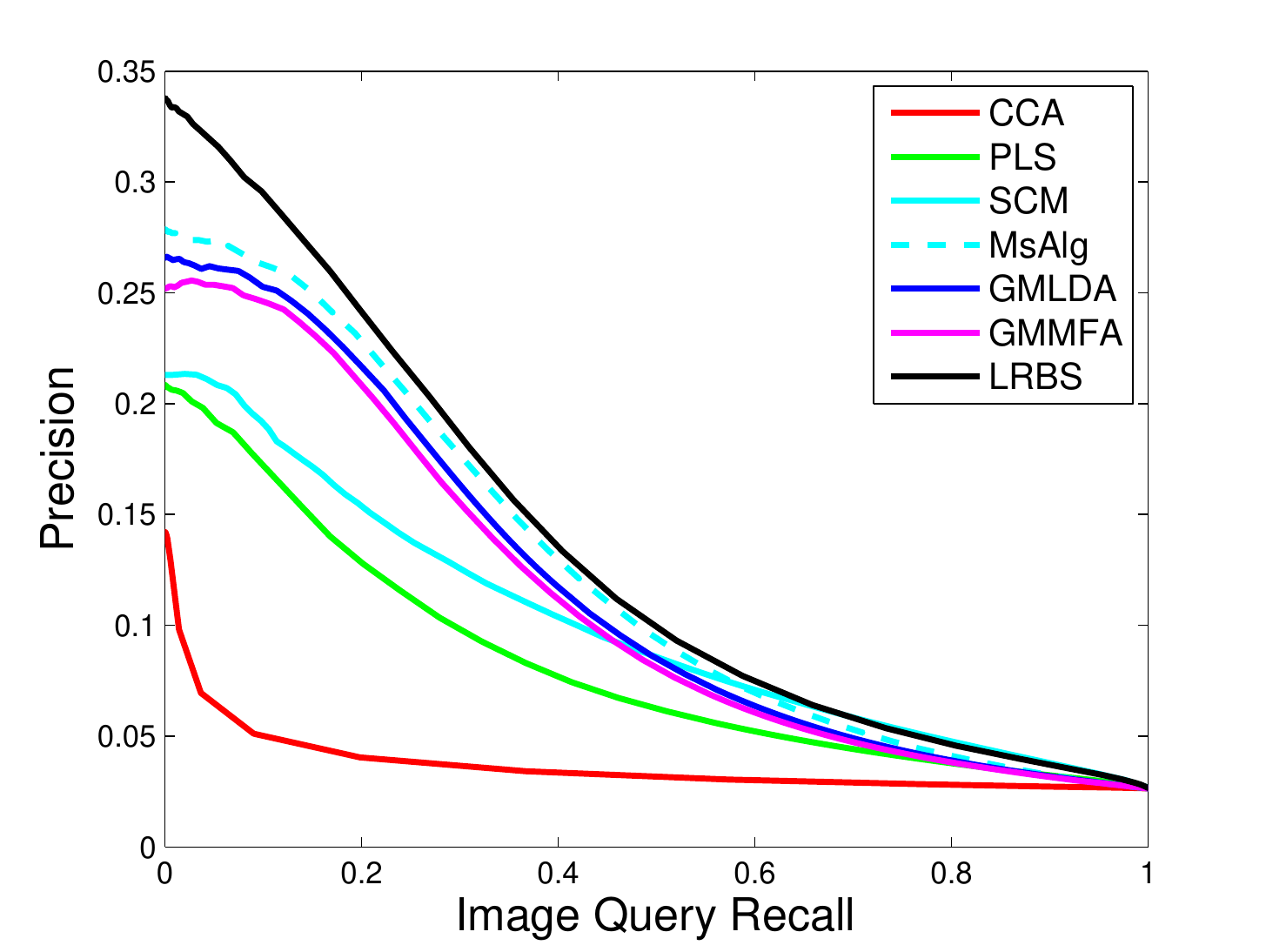}\hspace{12mm}
\includegraphics[width=2.50in]{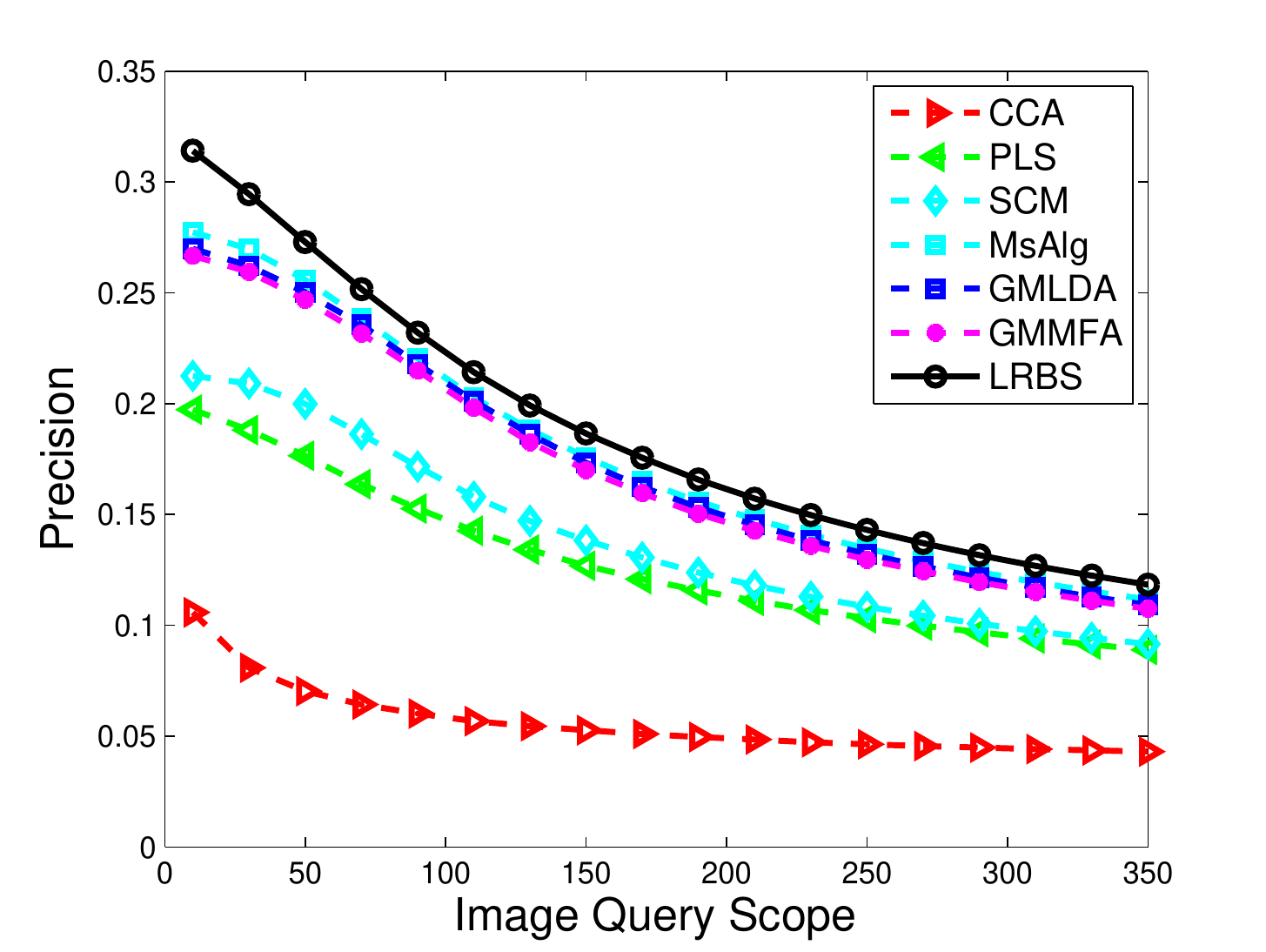} \\
(a) Image-to-Text \\
%\vspace{1ex}
\includegraphics[width=2.50in]{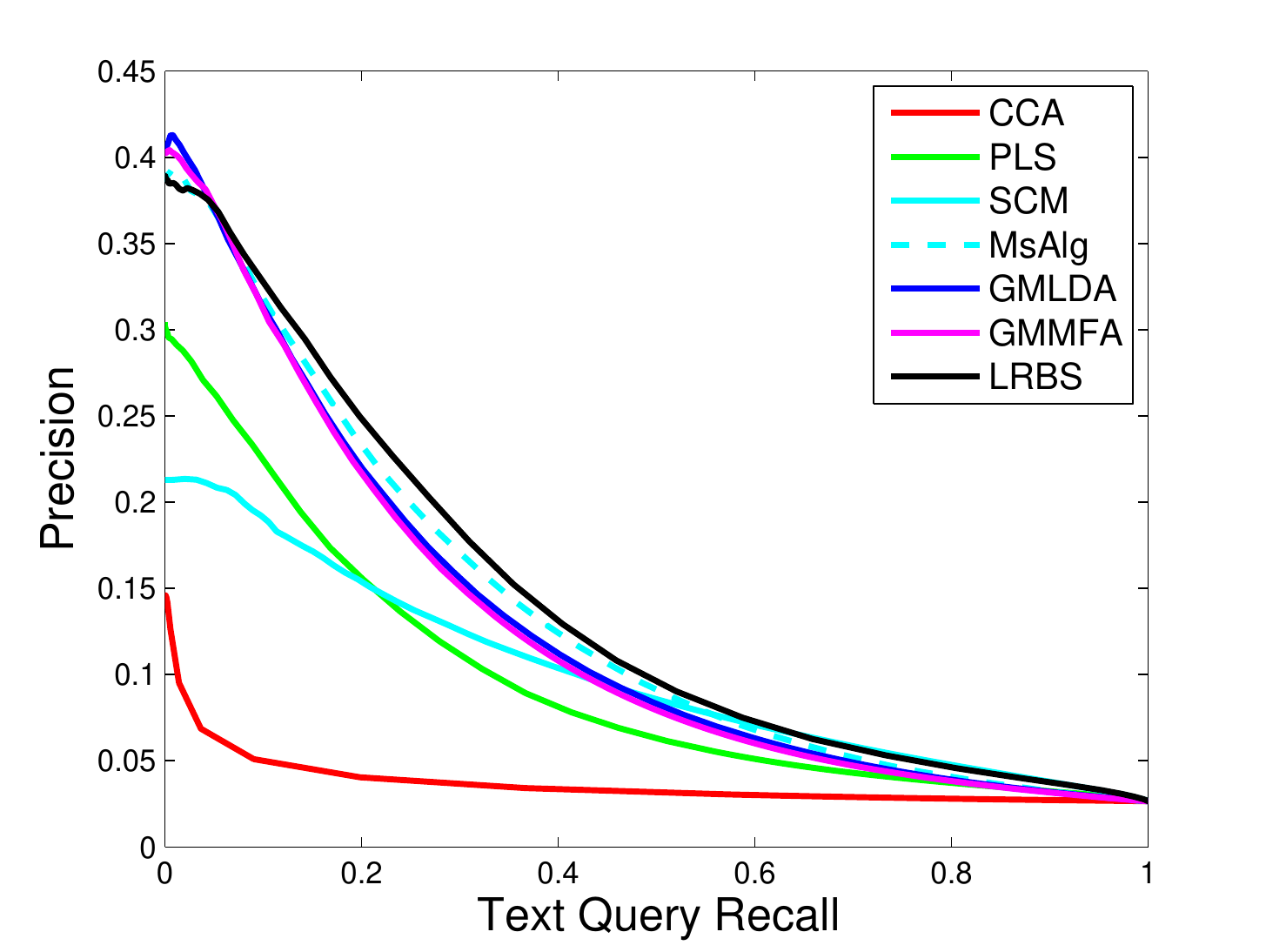} \hspace{12mm}
\includegraphics[width=2.50in]{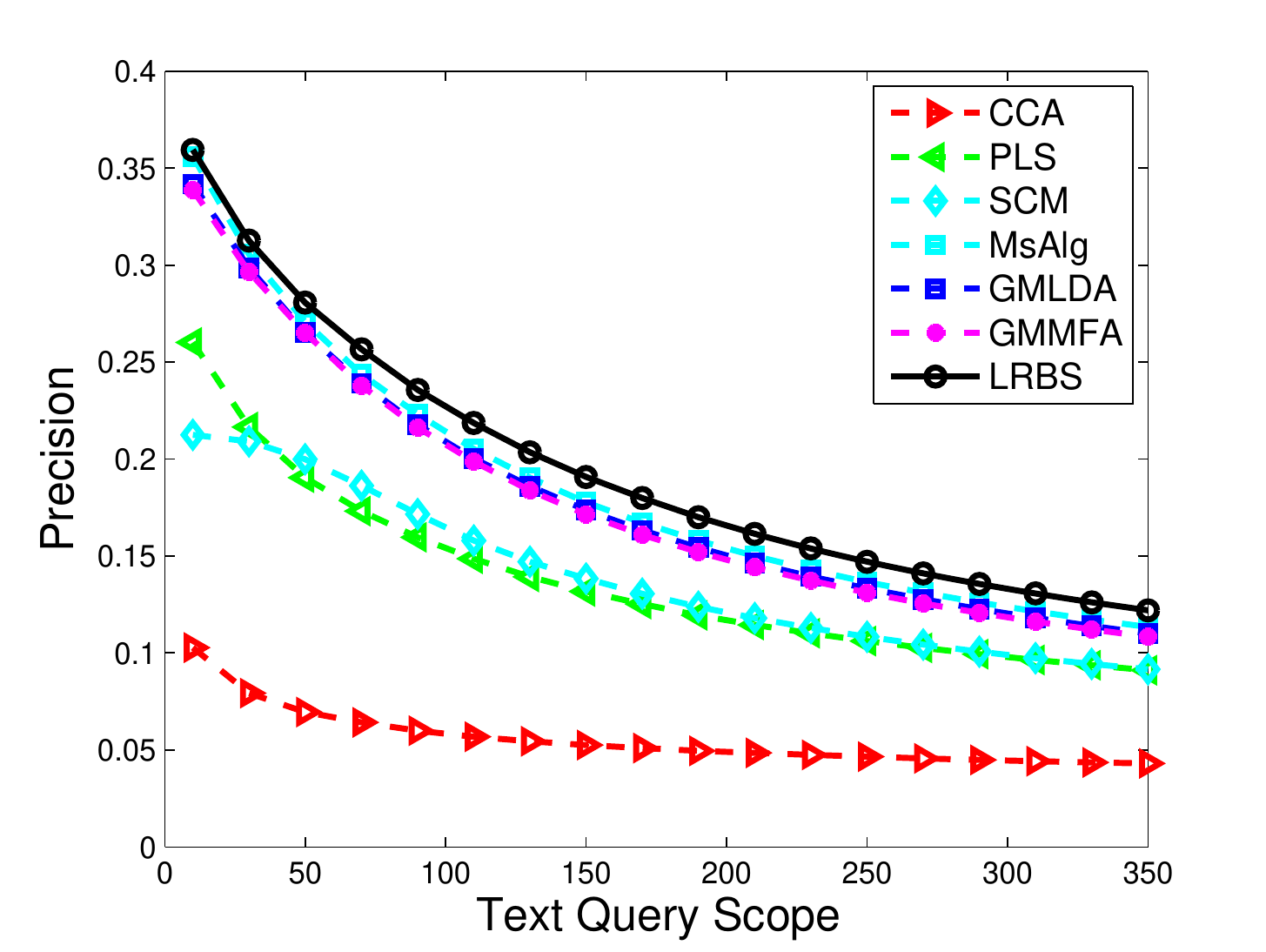}\\
(b) Text-to-Image
\caption{Precision-recall and precision-scope curves on the NUS-WIDE database for the Image-to-Text retrieval experiment and the Text-to-Image retrieval experiment.}
\label{fig:nuswide-rst}
\end{figure*}

\subsection{Databases}

In the experiments, three famous images and texts databases in the multimedia retrieval field are used to evaluate the performance of the proposed algorithm, namely the Pascal VOC2007 database, the Wikipedia database, and the NUS-WIDE database.

{\bf The Pascal VOC2007} dataset consists of 9963 images from 20 categories \cite{PASCAL-VOC2007}, which was split into a training set with 5011 images and a test set with 4952 images. Since some of the images are with multi-labels, the images containing only one object were selected in the experiments \cite{Sharma-cvpr-12}. As a result, there are 2808 images for the training set and 2841 images for the test set. For the features, the 399-dimensional word frequency feature \cite{hwang-bmvc-10} was used for the text, and the convolutional neural network (CNN) feature, which was trained on the ImageNet \cite{Krizhevsky-nips-12}, was used as the image feature. The CNN source code, namely Decaf \cite{donahue-decaf}, can be freely downloaded on the web for research purpose\footnote{http://daggerfs.com/}. In the experiments, only the outputs of the sixth layer were used as the image feature, resulting in 4096 dimensions.
%\footnote{https://github.com/UCB-ICSI-Vision-Group/decaf-release/wiki}

%
\begin{table}[t]
\center
\caption{\small{MAP (\%) results on the Wikipedia database.}}
\begin{tabular}{l|c|c|c}
\hline
                & \multicolumn{3}{c}{Without PCA}        \\  \hline
Methods         & Image query & Text query  & ~Average~      \\    \hline
CCA/SCM         & --           & --            & --        \\
PLS             & 30.54        & 28.05         & 29.30     \\
MsAlg           & 37.28        & 32.70         & 34.99     \\
GMLDA           & 24.93        & 18.18         & 21.55     \\
GMMFA           & 24.03        & 16.51         & 20.27     \\
LRBS            &\textbf{44.48} & \textbf{37.70} &\textbf{41.09} \\
\hline\hline
                & \multicolumn{3}{c}{With PCA}             \\ \hline
Methods         & Image query & Text query  & ~Average~      \\  \hline
CCA             & 19.70        & 17.84         & 18.77     \\
PLS             & 30.55        & 28.03         & 29.29     \\
SCM             & 37.13        & 28.23         & 32.68     \\
MsAlg           & 36.07        & 30.75         & 33.41     \\
GMLDA           & 36.77        & 29.71         & 33.24     \\
GMMFA           & 38.74        & 31.09         & 34.91     \\
LRBS            & {\color{blue}\textbf{44.41}}   & \textbf{37.70}   & {\color{blue}\textbf{41.06}}  \\
\hline
\end{tabular}
\label{tlb:wiki-rst}
\end{table}

{\bf The Wikipedia} dataset\footnote{http://www.svcl.ucsd.edu/projects/crossmodal/} is constructed from the Wikipedia's ``featured articles", which is a continually updated collection selected and reviewed by the Wikipedia's editors since 2009. Among the articles in the collection, Rasiwasia et al. built a dataset by selecting ten popular categories \cite{Rasiwasia-mm-10}. Each image in the dataset is associated with a section of at least 70 words. In total, the dataset consists of 2866 image-document pairs. In \cite{Rasiwasia-mm-10}, the dataset was randomly split into a training set of 2173 image-document pairs and a test set of the remaining 693 pairs. The text feature was derived from the Latent Dirichlet Allocation model (LDA) with 200 topics \cite{jose-pami-2014}. Similar as on the Pascal VOC2007 database, the CNN feature was also used for the images on this dataset.

\begin{table}[t]
\center
\caption{\small{MAP (\%) results on the NUS-WIDE database.}}
\begin{tabular}{ l|c|c|c}
\hline
                & \multicolumn{3}{c}{Without PCA}        \\  \hline
Methods         & Image query & Text query  & ~Average~  \\    \hline
CCA/SCM         &   --         &  --         & --         \\
PLS             & 13.66        & 13.45       & 13.55      \\
MsAlg           & 20.70        & 19.43       & 20.07      \\
GMLDA           & 10.37        &  8.47       &  9.42     \\
GMMFA           &  9.80        &  8.00       &  8.90     \\
LRBS            &\textbf{21.75} &\textbf{20.79} &\textbf{21.27} \\
\hline\hline
                & \multicolumn{3}{c}{With PCA}        \\ \hline
Methods         & Image query & Text query  & ~Average~  \\    \hline
CCA             &  4.75        &  4.64       &  4.70      \\
PLS             & 13.77        & 13.44       & 13.61      \\
SCM             & 16.44        & 14.07       & 15.26    \\
MsAlg           & 20.02        & 18.20       & 19.11    \\
GMLDA           & 20.31        & 18.48       & 19.39     \\
GMMFA           & 19.98        & 18.32       & 19.15     \\
LRBS         &{\color{blue}\textbf{21.73}} & 2{\color{blue}\textbf{0.73}} &{\color{blue}\textbf{21.23}}  \\
\hline
\end{tabular}
\label{tlb:nuswide-rst}
\end{table}

\begin{figure*}[!htb]
\center
\includegraphics[width=6.95in]{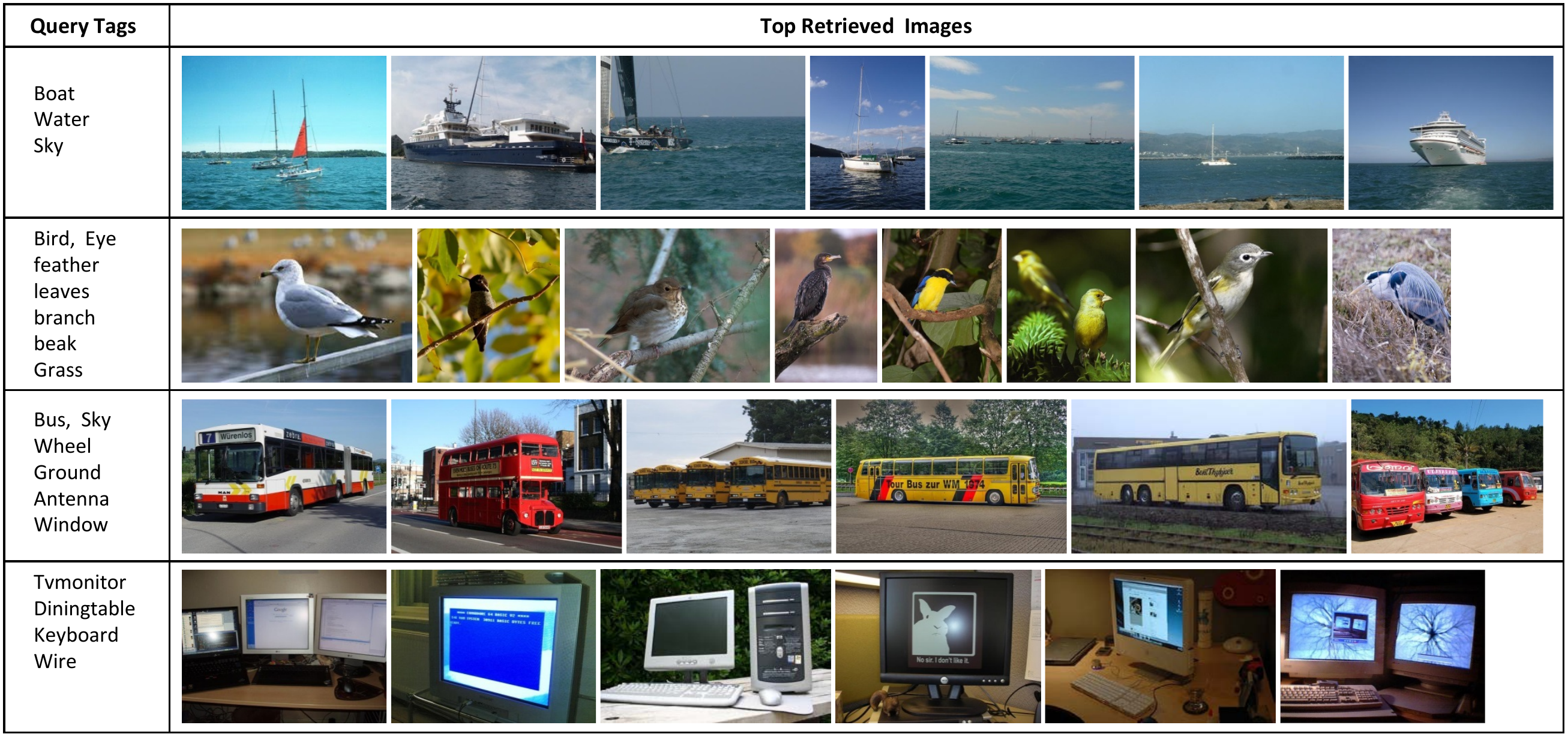}
\caption{The retrieved samples from texts to images by the proposed LRBS algorithm on the Pascal VOC2007 database.}
\label{fig:voc2007-retrie-sample}
\end{figure*}

{\bf The NUS-WIDE} dataset is a large-scale real-world database where the images were crawled from the ``Flickr'' website by the Media Search Lab in the National University of Singapore~\cite{nuswide-09}. The database is constructed with 269,648 images associated with tags. The organizers has labeled the images with 81 concepts for evaluation. They also provided all the URLs of the images on the web to facilitate further research\footnote{http://lms.comp.nus.edu.sg/research/NUS-WIDE.htm}. In the experiment, we used 40 concepts which contained the most uniquely labeled images, and 250 images per concept were randomly selected for evaluation. The selected images were further randomly separated into a training set and a test set, which contain 6000 images and 3630 images respectively. Again, the CNN feature for the images was used in the experiment. As for the text feature, the provided 1000-dimensional word frequency feature by the database organizers~\cite{nuswide-09} was imported.% In addition, the regularization parameter $\lambda$ of the proposed algorithm for the Pascal VOC2007 and the Wikipedia databases are 1e-6 and 1e-4, respectively.

\subsection{Experimental Results}

In the experiments, we found that the CCA, PLS and GMA algorithms all performed better with PCA dimensional reduction than without it. Thus, we displayed both the experimental results with PCA and without PCA for all the algorithms. With PCA, about 99\% energy were preserved for the features. As a result, the dimension of the CNN image feature was reduced to 1000 on the three databases. As for the text feature, it was reduced to 200 on the Pascal VOC2007 database, 180 on the Wikipedia and 800 on the NUS-WIDE database. The MAPs of the experimental results are shown in Table \ref{tlb:voc-rst}, Table \ref{tlb:wiki-rst} and Table \ref{tlb:nuswide-rst} for the VOC2007, the Wikipedia and the NUS-WIDE databases respectively, where the black bold numbers are the best performances, and the blue bold ones are the second best across the performances with PCA and not with PCA. The precision-recall and precision-scope curves are shown in Figure \ref{fig:voc-rst}, \ref{fig:wiki-rst} and Figure \ref{fig:nuswide-rst} for the three datasets, respectively.

From Table \ref{tlb:voc-rst}, Table \ref{tlb:wiki-rst} and Table \ref{tlb:nuswide-rst} it can be observed that the proposed LRBS algorithm achieves the best performance on the three databases. On the Wikipedia dataset, no matter with PCA and without PCA, the proposed method performs about 6\% higher than the second best algorithm (the GMMFA algorithm with PCA and the MsAlg algorithm without PCA). The followings are MsAlg with PCA, GMLDA with PCA and the SCM with PCA, which perform comparable to each other. This ranking is similar with that on the NUS-WIDE database. On the VOC2007 dataset, LRBS outperforms the second best one, the GMLDA algorithm with PCA, by 1\%. The followings are the SCM with PCA and the MsAlg algorithms. It can be also observed that the GMMFA and the GMLDA perform closely to each other on the three datasets.

It should be noted that the CCA without dimensional reduction cannot be applied since the feature matrix is not with full rank. This is the same with the SCM which needs CCA as the first step for correlation matching. Obviously, the GMA methods with PCA dimensional reduction perform better than without PCA. In contrast, for the PLS, MsAlg, and LRBS algorithms, they perform almost the same with PCA dimensional reduction and without it. The reason may lie in that the PCA dimension reduction removes part of the redundant information and noises. However, PCA may also discard some useful information at the same time. This may be the reason that the performances of the MsAlg and LRBS algorithms do not improve with PCA.

Table \ref{tlb:voc-rst}, \ref{tlb:wiki-rst} and \ref{tlb:nuswide-rst} also show that the supervised algorithms (the MsAlg, GMMFA, GMLDA, SCM and LRBS algorithms) clearly outperform unsupervised ones (the CCA and the PLS). It is because the class information used in supervised learning could partly reduce the semantic gap between low-level image features and high-level semantic meanings. The work in \cite{jose-pami-2014} also found that the combination of correlation matching and semantic matching can work better.

Figure \ref{fig:voc-rst}, Figure \ref{fig:wiki-rst} and Figure \ref{fig:nuswide-rst} display the precision-recall curves and the precision-scope curves of the compared algorithms for the image-text cross retrieval task. The performances of the GMA methods without PCA dimensional reduction are not shown in the figures due to the low performances. Note that the figures on show the best performance of each compared algorithm. For example, since the performance of the MsAlg method with PCA is slightly lower than without PCA, the figures just display the performance of the MsAlg without PCA. In these figures, it is also observed that the proposed algorithm has the best performance on all databases, followed by GMLDA, GMMFA and MsAlg methods, which are comparable to each other. From the precision-recall figures, it is clearly that with the same recall rate, the proposed method gets the highest precision in all compared algorithms, especially on the Wikipedia dataset. This is similar with the precision-scope curves, which plot the precision under different numbers of the top $r$ retrieved samples. In summary, the experimental results show that the proposed algorithm successfully learns a similarity function for cross-modal matching.
\begin{figure*}[!htb]
\center
\includegraphics[width=6.95in]{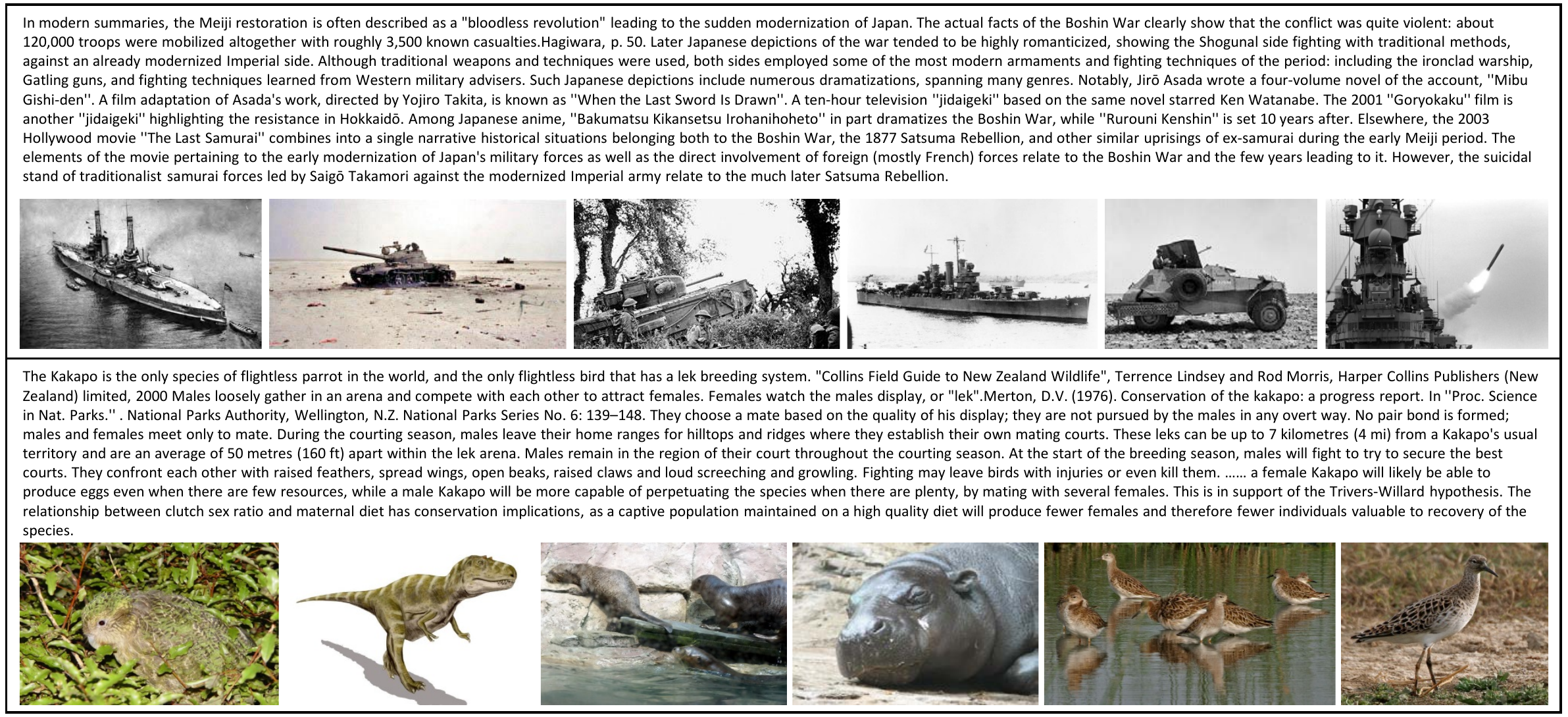}
\caption{The retrieved samples from texts to images by the proposed LRBS algorithm on the Wikipedia database. The figure shows the probe paragraphs and the corresponding top retrieved images.}
\label{fig:wiki2-retrie-sample}
\end{figure*}

\begin{figure}[t]
\center
\includegraphics[width=1.00in]{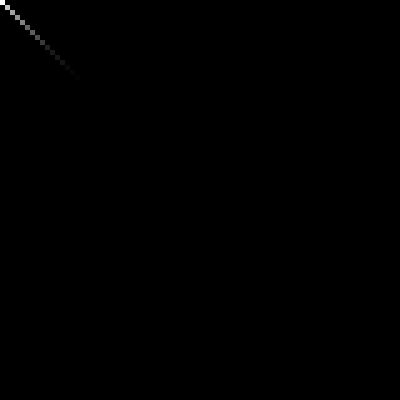}\hspace{1mm}
\includegraphics[width=1.00in]{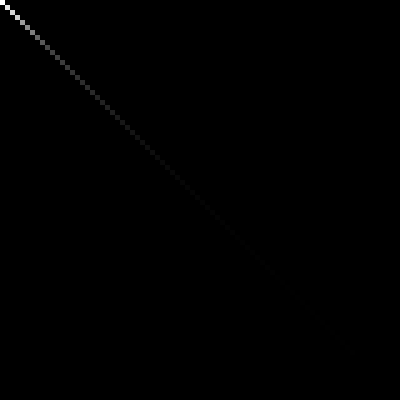}\hspace{1mm}
\includegraphics[width=1.00in]{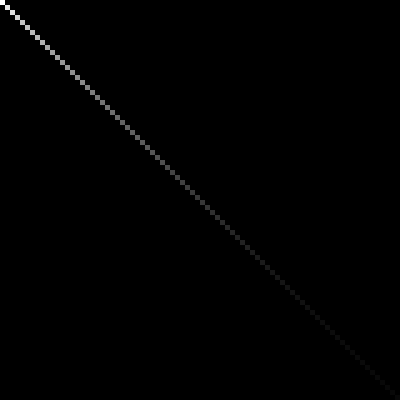}\\
(a) \hspace{22mm} (b) \hspace{22mm} (c) \\
\caption{The singular values of the learned similarity function matrix M on the three databases, where the values are shown in gray, and the 1 of gray value is the most lightest. (a) On the Wikipedia database; (b) On the Pascal VOC2007 database; (c) On the NUS-WIDE database.}
\label{fig:lowrank-rst}
\end{figure}

In Figure \ref{fig:lowrank-rst}, the singular values of the learned similarity metric matrix are shown, which contains the top 80 singular values. From the figure we can see that the learned $\mathbf{M}$ has a low rank, indicating that the proposed LRBS algorithm succeeds to find the similar structures in the data. It can be seen that the similarity metric of the Wikipeda dataset has the lowest rank. This is partly because the Wikipedia dataset only contains 10 categories. In contrast, the NUS-WIDE dataset contains the most categories, thus its similarity metric has a relatively high rank. In Figure \ref{fig:voc2007-retrie-sample} and Figure \ref{fig:wiki2-retrie-sample}, some retrieved samples by the proposed LRBS algorithm in the experiment are shown. The retrieved samples also indicate that the LRBS is an effective cross-modality information retrieval algorithm.

\section{Conclusion}
In this paper, we proposed a novel similarity function learning method inspired by metric learning algorithms. The similarity function is learned in a bilinear form of two different modality features, which can handle cross-modal features with different dimensions. The accelerated proximal gradient method is successfully imported to find the optimal solution with a fast convergence rate. Besides, we also imported the nuclear-norm to explore the structures and connections of the two modalities. The experiments are evaluated on three famous multimedia databases for the image-text cross-modal information retrieval problem, showing that the proposed algorithm has the best performance compared to the state-of-the-art algorithms.

%ACKNOWLEDGMENTS are optional
\section{Acknowledgments}
This work was supported in part by the National Basic Research Program of China (Grant 2012CB316304), the National Natural Science Foundation of China (Grants 61272331, 91338202 and 91438105), the Strategic Priority Research Program of the Chinese Academy of Sciences Grant (XDA06030200), Beijing Key Lab of Intelligent Telecommunication Software and Multimedia (ITSM201502), Guangxi Key Laboratory of Trusted Software (KX201418).

\bibliographystyle{abbrv}
\bibliography{aaai15ML-short}

\begin{thebibliography}{10}

\bibitem{bach-jmlr-08}
F.~R. Bach.
\newblock Consistency of trace norm minimization.
\newblock {\em Journal of Machine Learning Research}, 9:1019--1048, 2008.

\bibitem{Bar-icml-03}
A.~Bar-Hillel, T.~Hertz, N.~Shental, and D.~Weinshall.
\newblock Learning distance functions using equivalence relations.
\newblock In {\em ICML}, pages 11--18, 2003.

\bibitem{beck-siam-09}
A.~Beck and M.~Teboulle.
\newblock A fast iterative shrinkage-thresholding algorithm for linear inverse
  problems.
\newblock {\em SIAM J. Imaging Sciences}, 2(1):183--202, 2009.

\bibitem{Bronstein-cvpr-10}
M.~Bronstein, A.~Bronstein, F.~Michel, and N.~Paragios.
\newblock Data fusion through cross-modality metric learning using
  similarity-sensitive hashing.
\newblock In {\em CVPR}, pages 3594--3601, 2010.

\bibitem{cai-siam-10}
J.-F. Cai, E.~J. Cand\`{e}s, and Z.~Shen.
\newblock A singular value thresholding algorithm for matrix completion.
\newblock {\em SIAM Journal on Optimization}, 20(4):1956--1982, 2010.

\bibitem{gal-nips-09}
G.~Chechik, U.~Shalit, V.~Sharma, and S.~Bengio.
\newblock An online algorithm for large scale image similarity learning.
\newblock In {\em NIPS}, pages 306--314, 2009.

\bibitem{nuswide-09}
T.-S. Chua, J.~Tang, R.~Hong, H.~Li, Z.~Luo, and Y.-T. Zheng.
\newblock {NUS-WIDE: A Real-World Web Image Database from National University
  of Singapore}.
\newblock In {\em ACM Conf. on Image and Video Retrieval}, 2009.

\bibitem{jose-pami-2014}
J.~Costa~Pereira, E.~Coviello, G.~Doyle, N.~Rasiwasia, G.~Lanckriet, R.~Levy,
  and N.~Vasconcelos.
\newblock On the role of correlation and abstraction in cross-modal multimedia
  retrieval.
\newblock {\em TPAMI}, 36(3):521--535, 2014.

\bibitem{davis-icml-07}
J.~V. Davis, B.~Kulis, P.~Jain, S.~Sra, and I.~S. Dhillon.
\newblock Information-theoretic metric learning.
\newblock In {\em ICML}, volume 227, pages 209--216, 2007.

\bibitem{donahue-decaf}
J.~Donahue, Y.~Jia, O.~Vinyals, J.~Hoffman, N.~Zhang, E.~Tzeng, and T.~Darrell.
\newblock Decaf: A deep convolutional activation feature for generic visual
  recognition.
\newblock {\em arXiv preprint arXiv:1310.1531}, 2013.

\bibitem{Duan-ICML-2012}
L.~Duan, D.~Xu, and I.~W. Tsang.
\newblock Learning with augmented features for heterogeneous domain adaptation.
\newblock In {\em ICML}, pages 711--718, 2012.

\bibitem{PASCAL-VOC2007}
M.~Everingham, L.~Van~Gool, C.~K.~I. Williams, J.~Winn, and A.~Zisserman.
\newblock The {PASCAL} {V}isual {O}bject {C}lasses {C}hallenge 2007 {R}esults,
  2007.

\bibitem{gong-ijcv-14}
Y.~Gong, Q.~Ke, M.~Isard, and S.~Lazebnik.
\newblock A multi-view embedding space for modeling internet images, tags, and
  their semantics.
\newblock {\em International Journal of Computer Vision}, 106(2):210--233,
  2014.

\bibitem{guillaumin-iccv-09}
M.~Guillaumin, J.~J. Verbeek, and C.~Schmid.
\newblock Is that you? metric learning approaches for face identification.
\newblock In {\em ICCV}, pages 498--505, 2009.

\bibitem{Hardoon-neco-04}
D.~R. Hardoon, S.~Szedm\'{a}k, and J.~Shawe-Taylor.
\newblock Canonical correlation analysis: An overview with application to
  learning methods.
\newblock {\em Neural Computation}, 16(12):2639--2664, 2004.

\bibitem{hoi-cvpr-06}
S.~C.~H. Hoi, W.~Liu, M.~R. Lyu, and W.-Y. Ma.
\newblock Learning distance metrics with contextual constraints for image
  retrieval.
\newblock In {\em CVPR}, pages 2072--2078, 2006.

\bibitem{hwang-bmvc-10}
S.~J. Hwang and K.~Grauman.
\newblock Accounting for the relative importance of objects in image retrieval.
\newblock In {\em BMVC}, pages 1--12, 2010.

\bibitem{Ji-ICML-2009}
S.~Ji and J.~Ye.
\newblock An accelerated gradient method for trace norm minimization.
\newblock In {\em ICML}, pages 457--464, 2009.

\bibitem{jia-iccv-11}
Y.~Jia, M.~Salzmann, and T.~Darrell.
\newblock Learning cross-modality similarity for multinomial data.
\newblock In {\em ICCV}, pages 2407--2414, 2011.

\bibitem{Kan-eccv-12}
M.~Kan, S.~Shan, H.~Zhang, S.~Lao, and X.~Chen.
\newblock Multi-view discriminant analysis.
\newblock In {\em ECCV (1)}, volume 7572, pages 808--821, 2012.

\bibitem{Krizhevsky-nips-12}
A.~Krizhevsky, I.~Sutskever, and G.~E. Hinton.
\newblock Imagenet classification with deep convolutional neural networks.
\newblock In {\em NIPS}, pages 1106--1114, 2012.

\bibitem{kulis-cvpr-11}
B.~Kulis, K.~Saenko, and T.~Darrell.
\newblock What you saw is not what you get: Domain adaptation using asymmetric
  kernel transforms.
\newblock In {\em CVPR}, pages 1785--1792, 2011.

\bibitem{Lampert-eccv-10}
C.~H. Lampert and O.~Kr\"{o}mer.
\newblock Weakly-paired maximum covariance analysis for multimodal
  dimensionality reduction and transfer learning.
\newblock In {\em ECCV}, pages 566--579, 2010.

\bibitem{Li-iccv-11}
A.~Li, S.~Shan, X.~Chen, and W.~Gao.
\newblock Face recognition based on non-corresponding region matching.
\newblock In {\em International Conference on Computer Vision}, pages
  1060--1067, 2011.

\bibitem{mignon-accv-12}
A.~Mignon and F.~Jurie.
\newblock {CMML: a New Metric Learning Approach for Cross Modal Matching}.
\newblock In {\em {Asian Conference on Computer Vision}}, 2012.

\bibitem{Nes-book-03}
Y.~Nesterov.
\newblock {\em Introductory Lectures on Convex Optimization: {A Basic Course}}.
\newblock Kluwer Academic Publishers, 2003.

\bibitem{Ngiam-icml-11}
J.~Ngiam, A.~Khosla, M.~Kim, J.~Nam, H.~Lee, and A.~Y. Ng.
\newblock Multimodal deep learning.
\newblock In {\em ICML}, pages 689--696, 2011.

\bibitem{pan-sigir-14}
Y.~Pan, T.~Yao, T.~Mei, H.~Li, C.-W. Ngo, and Y.~Rui.
\newblock Click-through-based cross-view learning for image search.
\newblock In {\em ACM conference on Research and Development in Information
  Retrieval (SIGIR)}, 2014.

\bibitem{Rasiwasia-mm-10}
N.~Rasiwasia, J.~C. Pereira, E.~Coviello, G.~Doyle, G.~R.~G. Lanckriet,
  R.~Levy, and N.~Vasconcelos.
\newblock A new approach to cross-modal multimedia retrieval.
\newblock In {\em ACM Multimedia}, pages 251--260, 2010.

\bibitem{Rosipal-slsfs-06}
R.~Rosipal and N.~Kr\"{a}mer.
\newblock Overview and recent advances in partial least squares.
\newblock In {\em SLSFS}, pages 34--51. Springer, 2006.

\bibitem{Sharma-cvpr-11}
A.~Sharma and D.~W. Jacobs.
\newblock {Bypassing synthesis: PLS for face recognition with pose,
  low-resolution and sketch}.
\newblock In {\em CVPR}, pages 593--600, 2011.

\bibitem{Sharma-cvpr-12}
A.~Sharma, A.~Kumar, H.~D. III, and D.~W. Jacobs.
\newblock Generalized multiview analysis: A discriminative latent space.
\newblock In {\em CVPR}, pages 2160--2167, 2012.

\bibitem{socher-nips-13}
R.~Socher, M.~Ganjoo, C.~D. Manning, and A.~Y. Ng.
\newblock Zero-shot learning through cross-modal transfer.
\newblock In {\em NIPS}, pages 935--943, 2013.

\bibitem{Srivastava-nips-12}
N.~Srivastava and R.~Salakhutdinov.
\newblock Multimodal learning with deep boltzmann machines.
\newblock In {\em NIPS}, pages 2231--2239, 2012.

\bibitem{Tseng-SIAM-08}
P.~Tseng.
\newblock On accelerated proximal gradient methods for convex-concave
  optimization.
\newblock {\em submitted to SIAM Journal on Optimization}, 2008.

\bibitem{wang-iccv-13}
K.~Wang, R.~He, W.~Wang, L.~Wang, and T.~Tan.
\newblock Learning coupled feature spaces for cross-modal matching.
\newblock {\em International Conference on Computer Vision}, 2013.

\bibitem{weinberger-nips-06}
K.~Weinberger, J.~Blitzer, and L.~Saul.
\newblock Distance metric learning for large margin nearest neighbor
  classification.
\newblock {\em NIPS}, pages 1473--1480, 2006.

\bibitem{wu-report-10}
W.~Wu, J.~Xu, and H.~Li.
\newblock Learning similarity function between objects in heterogeneous spaces.
\newblock Technical Report MSR-TR-2010-86, 2010.

\bibitem{xing-nips-2002}
E.~P. Xing, A.~Y. Ng, M.~I. Jordan, and S.~J. Russell.
\newblock Distance metric learning with application to clustering with
  side-information.
\newblock In {\em NIPS}, pages 505--512, 2002.

\bibitem{zhai-aaai-13}
X.~Zhai, Y.~Peng, and J.~Xiao.
\newblock Heterogeneous metric learning with joint graph regularization for
  cross- media retrieval.
\newblock In {\em AAAI}, pages 1198--1204, 2013.

\bibitem{funZhu-cikm-14}
F.~Zhu, L.~Shao, and M.~Yu.
\newblock Cross-modality submodular dictionary learning for information
  retrieval.
\newblock In {\em CIKM}, pages 1479--1488, 2014.

\bibitem{zhuang-wsdm-11}
J.~Zhuang and S.~C.~H. Hoi.
\newblock A two-view learning approach for image tag ranking.
\newblock In {\em WSDM}, pages 625--634, 2011.

\bibitem{zhuang-aaai-13}
Y.~Zhuang, Y.~F. Wang, F.~Wu, Y.~Zhang, and W.~Lu.
\newblock Supervised coupled dictionary learning with group structures for
  multi-modal retrieval.
\newblock In {\em AAAI}, pages 1070--1076, 2013.

\end{thebibliography}

\appendix
\section{Proof of Theorem 1}

%------------------------------
\begin{proof}
Considering that the objective function in Eqn. (\ref{eqn:square-rank2}) is a strongly convex function, a unique solution exists. Thus we just need to prove that the optimal solution is equal to $C_{\gamma}(\mathbf{M})$ \cite{cai-siam-10}. Considering that $\mathbf{\hat{M}}$ is the optimal solution of Eqn.(\ref{eqn:square-rank2}) if and only if $\mathbf{0}$ is a subgradient of the function at the point $\mathbf{\hat{M}}$, we have
\begin{equation}
\centering
\mathbf{0} \in \mathbf{\hat{M}} - \mathbf{L} + \gamma \partial \|\mathbf{\hat{M}}\|_{*},
\end{equation}
where the $\partial \|\mathbf{\hat{M}}\|_{*}$ is the subgradient of the nuclear norm. Let the SVD decomposition of an arbitrary matrix $\mathbf{A}\in\mathds{R}^{m\times n}$ is  $\mathbf{A}=\mathbf{P}_1\mathbf{\Lambda}\mathbf{P}_2^T$, then the subgradient of its nuclear norm is \cite{bach-jmlr-08,cai-siam-10}
\begin{equation}
\centering
\begin{split}
\partial \|\mathbf{A}\|_{*} =\{& \mathbf{P}_1 \mathbf{P}_2^T +\mathbf{S} : \mathbf{S}\in\mathds{R}^{m\times n}, \mathbf{P}_1^T \mathbf{S}=0, \\
& \mathbf{S} \mathbf{P}_2=0, \|\mathbf{S}\|_2\leq 1 \}.
\end{split}
\end{equation}

Denote the SVD decomposition of $\mathbf{L}$ in Eqn.(\ref{eqn:square-rank2}) as
\begin{equation}
\centering
\mathbf{L} = \mathbf{U}_0\mathbf{\Sigma}_0\mathbf{V}_0^{T} + \mathbf{U}_1\mathbf{\Sigma}_1\mathbf{V}_1^{T}, \nonumber
\end{equation}
where $\mathbf{U}_0\mathbf{\Sigma}_0\mathbf{V}_0^{T}$ is the part of SVD with singular values greater than $\gamma$, and the $\mathbf{U}_0\mathbf{\Sigma}_0\mathbf{V}_0^{T}$ corresponds to the remaining part. By denoting $\mathbf{\hat{M}}=C_{\gamma}(\mathbf{M})$, we have
\begin{equation}
\centering
\mathbf{\hat{M}} = \mathbf{U}_0(\mathbf{\Sigma}_0-\gamma\mathbf{I})\mathbf{V}_0^{T}. \nonumber
\end{equation}
Therefore,
\begin{equation}
\centering
\mathbf{L} - \mathbf{\hat{M}} = \mathbf{L} - C_{\gamma}(\mathbf{M}) = \gamma(\mathbf{U}_0\mathbf{V}_0^{T}+\mathbf{S}), \nonumber
\end{equation}
where
\begin{equation}
\centering
\mathbf{S} = \gamma^{-1}\mathbf{U}_1\Sigma_1\mathbf{V}_1^{T}. \nonumber
\end{equation}
It turns out that $\mathbf{U}_0^T\mathbf{S}=0$, $\mathbf{S}\mathbf{V}_0=0$, and $\|\mathbf{S}\|_2\leq1$ since $\Sigma_1$ is bounded by $\gamma$. Finally, we have proved that $\mathbf{L} - C_{\gamma}(\mathbf{M}) \in \gamma\partial \|C_{\gamma}(\mathbf{M})\|_{*}$, which shows that $C_{\gamma}(\mathbf{M})$ is the optimal solution of Eqn.(\ref{eqn:square-rank2}).

\end{proof}

\end{document}